\documentclass[a4paper,fleqn]{cas-dc}

\usepackage[authoryear]{natbib}
\usepackage{subfigure}
\usepackage{graphicx}

\def\tsc#1{\csdef{#1}{\textsc{\lowercase{#1}}\xspace}}
\tsc{WGM}
\tsc{QE}
\tsc{EP}
\tsc{PMS}
\tsc{BEC}
\tsc{DE}
\begin{document}

\let\WriteBookmarks\relax
\def\floatpagepagefraction{1}
\def\textpagefraction{.001}
\shorttitle{Review of Single-cell RNA-seq Data Clustering}
\shortauthors{S. Zhang et~al.}

\title [mode = title]{Review of Single-cell RNA-seq Data Clustering for Cell Type Identification and Characterization}                      

\author[1]{Shixiong Zhang}
\ead{sxzhang7-c@my.cityu.edu.hk}
\credit{Conceptualization of this study, Writing - Original draft preparation}

\author[1]{Xiangtao Li}

\ead{lixt314@nenu.edu.cn}

\author[1]{Qiuzhen Lin}
\ead{qiuzhlin@szu.edu.cn}

\author[1]{Ka-Chun Wong}
\cormark[1]
\ead{kc.w@cityu.edu.hk}
\credit{Writing - Original draft preparation}

\address[1]{Department of Computer Science, City University of Hong Kong, Kowloon Tong, Hong Kong SAR.}

\cortext[cor1]{Corresponding author.}

\begin{abstract}
 In recent years, the advances in single-cell RNA-seq techniques have enabled us to perform large-scale transcriptomic profiling at single-cell resolution in a high-throughput manner. 
 Unsupervised learning such as data clustering has become the central component to identify and characterize novel cell types and gene expression patterns.
  
 In this study, we review the existing single-cell RNA-seq data clustering methods with critical insights into the related advantages and limitations. In addition, we also review the upstream single-cell RNA-seq data processing techniques such as quality control, normalization, and dimension reduction. 
 We conduct performance comparison experiments to evaluate several popular single-cell RNA-seq clustering approaches on two single-cell transcriptomic datasets.
 
\end{abstract}

\begin{keywords}
Single-cell RNA-seq \sep Clustering \sep Cell types \sep Identification \sep Characterization \sep Review
\end{keywords}

\maketitle

\section{Introduction}

With the unabated progress in high-throughput sequencing technologies, single-cell RNA-seq has become a powerful approach to simultaneously measure cell-to-cell expression variability of thousands or even hundreds of thousands of genes \citep{Grun2015Single-cellTypes, Shapiro2013Single-cellScience} at single cell resolution. Such high-throughput transcriptomic profiling can capture the gene transcriptional activities to reveal cell identities and functions \citep{Kiselev2019ChallengesData, Patel2014Single-cellGlioblastoma} and discover cell types \citep{Shalek2014Single-cellVariation, Xu2015IdentificationMethod, Zeisel2015BrainRNA-seq} or even rare cell types \citep{vanUnen2017VisualTypes, Jiang2016GiniClust:Index, Grun2015Single-cellTypes}. Hence, one of the most common goals of those single-cell studies is to identify cell subpopulations under different contexts \citep{Yang2017SAIC:Data}. The gene expression patterns of those subpopulations help us distinguish various cell types and functions, identifying different cell types.

Diverse computational approaches based on data clustering have emerged to interpret and understand single-cell RNA-seq data \citep{Jiang2016GiniClust:Index, Lin2017CIDR:Data, Yang2017SAIC:Data, Wolf2018SCANPY:Analysis, Zheng2019SinNLRR}. The advances in single-cell clustering has also initiated the development of multiple atlas projects such as Mouse Cell Atlas \citep{Han2018MappingMicrowell-Seq}, Aging Drosophila Brain Atlas \citep{Davie2018ABrain}, and Human Cell Atlas \citep{Rozenblatt-Rosen2017TheReality}. However, several technical challenges are still involved in single-cell RNA-seq clustering. Low-quality cells/genes, amplification biases, and other confounding factors can affect the downstream clustering performance. In addition, given the whole transcriptome range of RNA-seq the curse of dimensionality should be expected \citep{Andrews2018IdentifyingScRNASeq}. Thus the data preprocessing steps including quality control, normalization, and dimensional reduction have become necessary before downstream interpretation. In addition, the tissue heterogeneity can also affect the single-cell RNA-seq clustering performance to detect rare cell types  \citep{vanUnen2017VisualTypes, Jiang2016GiniClust:Index, Grun2015Single-cellTypes}.

\begin{figure}
	\centering
	\includegraphics[width=0.45\textwidth]{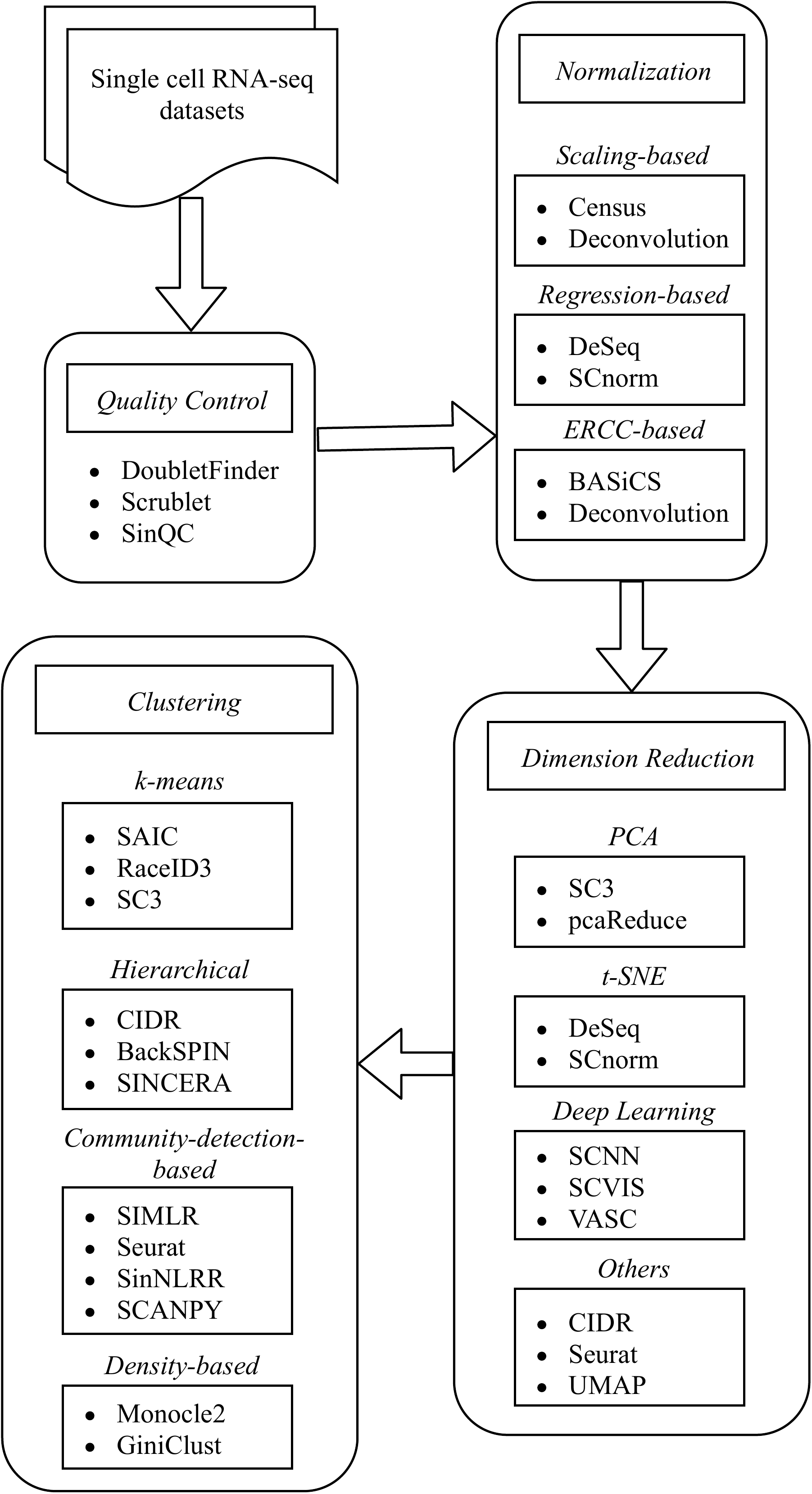}
    	\caption{Workflow of single cell RNA-seq data clustering.}
	\label{FIG:1}
\end{figure}

In this study, we review the recently developed computational clustering approaches for understanding and interpreting single-cell RNA-seq data. We also review the upstream single-cell RNA-seq data preprocessing steps such as quality control, row/column normalization, and dimension reduction before clustering is performed. Four roughly-classified categories of single-cell RNA-seq clustering methods and its application are discussed in terms of the strengths and limitations, including {\it k}-means clustering, hierarchical clustering, community-detection-based clustering, and density-based clustering. Figure~\ref{FIG:1} depicts the workflow of single cell RNA-seq data clustering by data processing (quality control, normalization, and dimension reduction) and clustering methods. The strengths and limitations are discussed in following sections to guide selection of different tools. In addition, we conduct several experiments on single-cell RNA-seq datasets to evaluate and compare those clustering methods. 

\section{Data preprocessing}
   
Given the technical variations and noises, data preprocessing is essential for unsupervised cluster analysis on single-cell RNA-seq data. Quality control is performed to remove the low-quality transcriptomic profile due to capture inefficiency; the single-cell RNA-seq reads should be normalized to remove any amplification biase, sample variation, and other technical confounding factors; dimensional reduction is conducted to project the high-dimensional single-cell RNA-seq data into low-dimensional space. Those upstream steps could have substantial impacts on downstream tasks. Therefore, a myriad of tools have been developed to address the above issues.

\subsection{Quality control}

Quality control (QC) aims to removing the unreliable cells or genes and other possible missing values for downstream interpretation \citep{Kiselev2019ChallengesData}. The technical reason for the presence of a large number of cells/genes can be attributed to the doublets with two or more cells suspended in one droplet; on the contrary, a small number of transcripts/genes may result from capture inefficiency (e.g., cell death, cell breakage, and a high fraction of mitochondrial counts) \citep{Grun2015DesignExperiments.,Lun2016AData}. In this section, we review several state-of-the-art tools or methods in assessing the raw reads and expression matrices of single-cell RNA-seq data. DoubletFinder \citep{McGinnis2019DoubletFinder:Neighbors} identifies doublets using gene expression features that significantly improves differential expression analysis performance. Scrublet avoids the need for expert knowledge or cell clustering by simulating multiplets from the data and building a nearest neighbor classifier \citep{Wolock2019Scrublet:Data}.  SinQC \citep{Jiang2016QualitySinQC} enables us to detect poor data quality, e.g. low mapped reads, a high fraction of mitochondrial counts or low library complexity.

\subsection{Normalization}

Technical artifacts or experimental noises (e.g. batch effect, insufficient counts, and zero inflation) of high-throughput transcriptomic sequencing may result in differences in expression measurements between samples (e.g. genes) \citep{Cole2019PerformanceRNA-Seq}. Several studies have revealed that those obvious differences can have a large impact on clustering \citep{Haghverdi2018BatchNeighbors, Butler2018IntegratingSpecies, Finak2015MAST:Data, Kharchenko2014BayesianAnalysis}. Therefore, normalization is essential for adjusting the differences in expression levels across different samples, replicates, or even batches.

The state-of-the-arts normalization methods have been developed for addressing those issues. We review three kinds of normalization methods as follow: 1) Scaling methods. \citet{L.Lun2016PoolingCounts} proposed a strategy to normalize single-cell RNA-seq data with zero counts. Census \citep{Qiu2017Single-cellCensus} converts conventional per-cell measures of relative expression values to transcript counts without the need for any spike-in standard or unique molecular identifiers, eliminating much of the apparent technical variability in single-cell experiments; 2) Regression-based methods. DESeq proposed by \cite{Anders2010DifferentialData} adopts local regression to link the variance and mean of negative binomial distribution over the observed counts, resulting in balanced differentially expressed genes. SCnorm \citep{Bacher2017SCnorm:Data} uses quantile regression to estimate the dependence of transcript expression on sequencing depth and scale factors to provide normalized expression estimates; 3) Methods based on spike-in External RNA Control Cortium (ERCC). \citet{Ding2015NormalizationExperiments} presented a normalization tool to remove technical noises and compute for the true gene expression levels based on spike-in ERCC. BASiCS \citep{Vallejos2015BASiCS:Data} can identify and remove the high and low levels of technical noises (counts). In addition to the above methods, the very simple and commonly used method is to transform read counts using logarithm with a pseudocount such as one \citep{Xu2015IdentificationMethod, Lin2017CIDR:Data, Butler2018IntegratingSpecies}. 

However, those normalization methods also suffer from limitations caused by the diverse assumptions and experimental protocols. The scaling methods cannot account for individual batch effects; the regression-based methods are sensitive to batch effects; ERCC-based methods are not suitable for endogenous and spiked-in transcripts \citep{Risso2014NormalizationSamples, Vallejos2017NormalizingOpportunities, Cole2019PerformanceRNA-Seq}.
 
\subsection{Dimension reduction}

Recent advances in single-cell RNA-seq have contributed to measure large-scale expression datasets with hundreds of thousands of genes while it also brings both opportunities and challenges in data analysis. Such high-dimensional gene expression data is unprecedentedly rich and should be well-explored. However, the past clustering methods may be unable to process and interpret such large-scale data. Hence, it is necessary to project the high-dimensional data to a lower-dimensional space using dimension reduction that can improve and refine the clustering results. In this section, we review several commonly used dimension reduction methods including principal component analysis, t-distributed stochastic neighbor embedding algorithm, deep learning models, and others.
 
\subsubsection{PCA}

Principal Component Analysis (PCA) is a typical linear projection method that projects a set of possibly correlated variables into a set of linearly orthogonal variables (principal components). Due to its conceptual simplicity and efficiency, PCA has been widely used in single-cell RNA-seq processing \citep{Jiang2016GiniClust:Index,Buettner2015ComputationalCells,Shalek2014Single-cellVariation,Usoskin2015UnbiasedSequencing,zurauskiene2016PcaReduce:Profiles,Kiselev2017SC3:Data}. Notably, SC3 \citep{Kiselev2017SC3:Data} applied PCA to transform the distance matrices as the input of consensus clustering; \citet{Shalek2014Single-cellVariation} used PCA for single-cell RNA-seq data spanning several experimental conditions. In addition, some extended and improved PCA-based methods have been developed including pcaReduce \citep{zurauskiene2016PcaReduce:Profiles} which applied PCA iteratively to provide low-dimensional principal component representations; \citet{Usoskin2015UnbiasedSequencing} proposed an unbiased iterative PCA-based process to identify distinct large-scale expression data patterns. However, PCA cannot capture the nonliner relationships between cells because of the high levels of dropout and noise \citep{Kiselev2019ChallengesData}. 

\subsubsection{t-SNE}

t-distributed Stochastic Neighbor Embedding (t-SNE) is the most commonly used nonlinear dimension reduction method which can uncover the relationships between cells. t-SNE converts data point similarity into probability and minimizes Kullback-Leibler divergence by gradient descent until convergence. In single-cell RNA-seq data analysis, t-SNE has become a cornerstone of dimension reduction and visualization for high-dimensional single-cell RNA-seq data \citep{Linderman2019FastData, Lin2017CIDR:Data, Butler2018IntegratingSpecies, Haghverdi2018BatchNeighbors, Ntranos2016FastCounts, Prabhakaran16, Zeisel2015BrainRNA-seq, Zhang2018AnDatasets, Li2017ReferenceTumors}. Especially, \citet{Linderman2019FastData} developed a fast interpolation-based t-SNE that dramatically accelerates the processing and visualization of rare cell populations for large datasets. Nonetheless, the limitations of t-SNE include the loss function is non-convex which can lead to different local optimality; the parameters in t-SNE are required to be tuned.

\subsubsection{Deep learning models}

In recent years, deep learning models (neural networks and variational auto-encoders) have shown superior performance in interpenetrating complex high-dimensional data. SCNN \citep{Lin2017UsingData} tested various neural networks architectures and incorporated prior biological knowledge to obtain the reduced dimension representation of single cell expression data. SCVIS \citep{Ding2018InterpretableModels} and VASC \citep{Wang2018VASC:Autoencoder} are both based on variational auto-encoders which can capture nonlinear relationships between cells and visualize the low-dimensional embedding in single-cell gene expression data. Up to now, those methods demonstrated superior ability of interpretation and compatibility on high-dimensional single-cell RNA-seq data.

\subsubsection{Other methods}

In addition, there are also other dimensional reduction methods such as CIDR \citep{Lin2017CIDR:Data}
applied principal coordinate analysis that preserves the distance information in low-dimension space from its high-dimension space; Seurat \citep{Butler2018IntegratingSpecies} is a toolkit for analysis of single cell RNA sequencing data and provides many dimension reduction methods such as PCA and t-SNE. Uniform Manifold Approximation and Projection (UMAP) \citep{Mcinnes2018UMAP:Archive} is a widely used technique for dimension reduction. UMAP provides increased speed and better preservation of data global structure for high dimensional datasets. It has been verified that it outperforms t-SNE \citep{Becht2019DimensionalityUMAP}.

\section{Clustering methods for single-cell RNA-seq}

Diverse types of clustering methods have been developed for detecting cell types from single-cell RNA-seq data. Those methods can be roughly classified into four categories including {\it k}-means clustering, hierarchical clustering, community-detection-based clustering, and density-based clustering. We review several computational applications of those clustering methods with their strengths and limitations. Table~\ref{tbl1} illustrates the overview of the state-of-the-arts clustering methods on single-cell RNA-seq data.

\begin{table*}[width=2.05\linewidth, pos=h]
	\caption{Overview of the state-of-the-arts clustering methods on single-cell RNA-seq data.}\label{tbl1}
	\begin{tabular*}{\tblwidth}{@{} LLLLLL@{} }
		\toprule
		Method                                     & Type      & \tabincell{L}{Open\\Source} & \tabincell{L}{Web\\ Sever}  & Strengths & Limitations\\
		\midrule
		
		\tabincell{L}{SAIC \citep{Yang2017SAIC:Data}}             & $k$-means & \checkmark  & $\times$   & \tabincell{L}{Low complexity;\\ Scalable to large data}& \tabincell{L}{Sensitive to outliers; No \\estimation of number of\\ clusters}  \\\hline
		\tabincell{L}{RaceID \citep{Grun2015Single-cellTypes},\\ RaceID2 \citep{Grun2016DeData},\\ RaceID3 \citep{Herman2018FateIDData}}    & $k$-means & \checkmark  & $\times$   & \tabincell{L}{Sensitive to rare cell types; \\Estimation of number of \\clusters}& \tabincell{L}{Not suitable for \\no rare cell types} \\\hline
		\tabincell{L}{pcaReduce\\ \citep{zurauskiene2016PcaReduce:Profiles}} & $k$-means/hierarchical & \checkmark & $\times$ & Hierarchy solutions& Not stable \\\hline
		\tabincell{L}{SC3 \citep{Kiselev2017SC3:Data}}            & $k$-means/hierarchical & \checkmark & $\times$ & \tabincell{L}{High accuracy; Estimation\\ of number of clusters\\ }& \tabincell{L}{High complexity; Not\\ scalable to large data} \\\hline
		\tabincell{L}{CIDR \citep{Lin2017CIDR:Data}}              & Hierarchical & \checkmark & $\times$ & Sensitive to Dropout& High complexity \\\hline
		\tabincell{L}{BackSPIN \citep{Zeisel2015BrainRNA-seq}}    & Hierarchical & \checkmark & $\times$ & \tabincell{L}{Simultaneously cluster\\ genes and cells}& High complexity \\	\hline	
		\tabincell{L}{SNN-Clip \citep{Xu2015IdentificationMethod}}& Cliques      & \checkmark & $\times$ & \tabincell{L}{Provide estimation of \\ number of clusters} & Non-scalable \\\hline
		\tabincell{L}{SIMLR \citep{Wang2017VisualizationLearning}}& Spectral     & \checkmark & $\times$ & \tabincell{L}{Suitable for data with\\ heterogeneity and noise}& \tabincell{L}{No estimation of \\ number of clusters} \\\hline
		\tabincell{L}{SinNLRR \citep{Zheng2019SinNLRR}}           & Spectral     & \checkmark & $\times$ & \tabincell{L}{Suitable for noise data} & \tabincell{L}{No estimation of \\ number of clusters} \\\hline
		\tabincell{L}{SCANPY \citep{Wolf2018SCANPY:Analysis}}     & Louvain & \checkmark  & $\times$   & \tabincell{L}{Low complexity\\ Scalable to large data} & \tabincell{L}{May not find\\ small community} \\\hline
		\tabincell{L}{Seurat \citep{Satija2015SpatialData}}       & Louvain & \checkmark  & $\times$   & \tabincell{L}{Low complexity\\ Scalable to large data}& \tabincell{L}{May not find\\ small community} \\\hline
		\tabincell{L}{GiniClust \citep{Jiang2016GiniClust:Index}} & Density-based & \checkmark  & $\times$   & \tabincell{L}{Available for detection\\ of rare cell types}& \tabincell{L}{Not sensitive to\\ large clusters} \\
		\bottomrule
	\end{tabular*}
\end{table*}

\subsection{$k$-means clustering}

$k$-means clustering is the most popular clustering approach, which iteratively finds a predefined number of $k$ cluster centers (centroids) by minimizing the sum of the squared Euclidean distance between each cell and its closest centroid. In addition, it is suitable for large datasets since it can scale linearly with the number of data points \citep{Lloyd1982LeastPCM}. 
 
 Several clustering tools based on $k$-means have been developed for interpreting single-cell RNA-seq data. SAIC \citep{Yang2017SAIC:Data} utilized an iterative $k$-means clustering to identify the optimal subset of signature genes that separate single cells into distinct clusters. pcaReduce \citep{zurauskiene2016PcaReduce:Profiles} is a hierarchical clustering method while it relies on $k$-means results as the initial clusters. RaceID \citep{Grun2015Single-cellTypes} applied $k$-means to unravel the heterogeneity of rare intestinal cell types \citep{Tibshirani2001EstimatingStatistic}.
 
 However, $k$-means clustering is an greedy algorithm that may fail to find its global optimum; the predefined number of clusters $k$ can affect the clustering results; and another disadvantage is its sensitivity to outliers since it tends to identify globular clusters, resulting in the failures in detecting of rare cell types. 
 
 To overcome the above drawbacks, SC3 \citep{Kiselev2017SC3:Data} integrated individual $k$-means clustering results with different initial conditions as the consensus clusters. RaceID2 \citep{Grun2016DeData} replaced the k-means clustering with k-medoids clustering that use 1- pearson's correlation instead of Euclidean distance as the clustering distance metric. RaceID3 \citep{Herman2018FateIDData}, as the advanced version of RaceID2 added feature selection and introduced random forest to reclassify $k$-means clustering results.

\subsection{Hierarchical clustering}

Hierarchical clustering is another widely used clustering algorithm on single-cell RNA-seq data. There are two types of hierarchical strategies including: 1) agglomerative clustering, the individual cells are progressively merged into clusters according to distance measures; 2) divisive clustering, each cluster is split into small groups recursively until individual data level. These two strategies build a hierarchical structure among the cells/genes and enable the improvement in finding rare cell types as small clusters. Hierarchical clustering does not require pre-determining the number of clusters and make assumptions for the distributions of single-cell RNA-seq data. Hence, many single-cell RNA-seq clustering methods have adopted it as part of the computational component.

CIDR \citep{Lin2017CIDR:Data} integrates both dimension reduction and clustering based on hierarchical clustering into single-cell RNA-seq analysis and uses implicit imputation process for dropout effects; it provides a stable estimation of pairwise cells distances. BackSPIN \citep{Zeisel2015BrainRNA-seq} developed a biclustering method based on divisive hierarchical clustering and sorting points into neighborhoods (SPIN) \citep{Tsafrir2005SortingMatrices} to simultaneously cluster genes and cells. The number of splits need to be set manually in BackSPIN. Although intensive splits can improve the clustering resolution, it is prone to over-partition. pcaReduce \citep{zurauskiene2016PcaReduce:Profiles} is an agglomerative hierarchical clustering approach with PCA which provides clustering results in a hierarchical. SINCERA \citep{Guo2015SINCERA} as a simple pipeline adopted hierarchical clustering with centered Pearson’s correlation and average linkage method to identify cell types.

The agglomerative hierarchical clustering has a time complexity of $ \mathcal{O}(N^3) $ while divisive clustering is $ \mathcal{O}(2^N) $. Although hierarchical clustering can reveal the hierarchical relations among cells/genes and does not require setting the number of clusters, it has high time complexity.

\subsection{Community-detection-based clustering}

Given the limitations of $k$-means and hierarchical clustering methods in large-scale datasets, community-detection-based clustering has been increasingly popular recently. Community detection is crucial in sociology, biology, and other systems that can be represented as graphs with nodes and edges. For single-cell RNA-seq data, nodes refer to cells and edge weights are represented by cell-cell pairwise distances. The idea of graph-based clustering is to delete the branch with maximum weights (cell-cell pairwise distances) in a dense graph (cell relationship network). There are three commonly used approaches for community-detection-based (graph-based) clustering including clique algorithm, spectral clustering, and Louvain algorithm \citep{Blondel2008FastNetworks}. 

A clique is a set of points fully connected to each other in a graph and represents a cluster (community). Although finding cliques in a graph is NP-complete, some studies have been conducted to address it such as heuristic optimization. SNN-Clip \citep{Xu2015IdentificationMethod} was proposed to leverage the concept of shared nearest neighbor to calculate cell similarity \citep{Zhang2009Genome} for finding all quasi-cliques since the shared nearest neighbor graph is sparse. SNN-Clip does not require specifying the number of clusters manually while it is non-scalable and the resultant clusters are not stable.

Spectral clustering is a widely used clustering method recently. It is designed to be adaptive to data distribution  by relying on the eigenvalues of cell similarity matrix. Nonetheless, the spectral clustering's time complexity is $ \mathcal{O}(N^3) $. 
SIMLR \citep{Wang2017VisualizationLearning} is an analytic framework for dimension reduction, clustering, and visualization of single-cell RNA-seq data. It is a method specificially designed at single-cell RNA-seq. SIMLR combines spectral clustering with multiple kernel similarity measures for clustering expression data generated from cross-platform and cross-condition experiments. In addition, SIMLR has an advantage in processing large-scale datasets with heavy noises. SinNLRR \citep{Zheng2019SinNLRR} was proposed to impose a non-negative and low rank structure on cell similarity matrix and then applied spectral clustering to detect cell types. However, the spectral clustering requires users to set the number of clusters in the data.

Louvain \citep{Blondel2008FastNetworks} is the most popular community detection algorithm and widely used to single-cell RNA-seq data. The time complexity of Louvain is $ \mathcal{O}(N\log(N)) $ which is lower than other community-detection-based algorithms. 
SCANPY \citep{Wolf2018SCANPY:Analysis} is a scalable toolkit for single-cell RNA-seq analysis and its clustering section is based on Louvain algorithm. SCANPY has advantages in scaling its computation with the number of cells (over one million). 
Seurat \citep{Satija2015SpatialData} also applied Louvain algorithm to cluster the cell types for the mapping of cellular localization.

\subsection{Density-based clustering} 

Density-based clustering methods separate data space into highly dense clusters. It can learn clusters with arbitrary shapes and identify noises (outliers). The most popular density-based clustering algorithm is DBSCAN \citep{Ester1996}. DBSCAN does not need to predetermine the number of clusters and its time complexity is $ \mathcal{O}(N\log(N)) $. However, DBSCAN requires user to set two parameters including $\epsilon$ (eps) and the minimum number of points required to form a dense region (minPts) \citep{Ester1996} that will affect its clustering results. \citet{Jiang2016GiniClust:Index} developed GiniClust, detecting rare cell types from single-cell gene expression data and its clustering method is based on DBSCAN. GiniClust is effective in finding rare cell types since it can be adaptively adjusted to set a lower $\epsilon$ . However, such a design may lead to unreasonable large cell clusters. Monocle2 \citep{Qiu2017ReversedTrajectories} also applied DBSCAN to identify the differential expressed genes between cells.

\section{Experimental evaluations for clustering methods}

\begin{figure}
	\centering
	\includegraphics[width=0.5\textwidth]{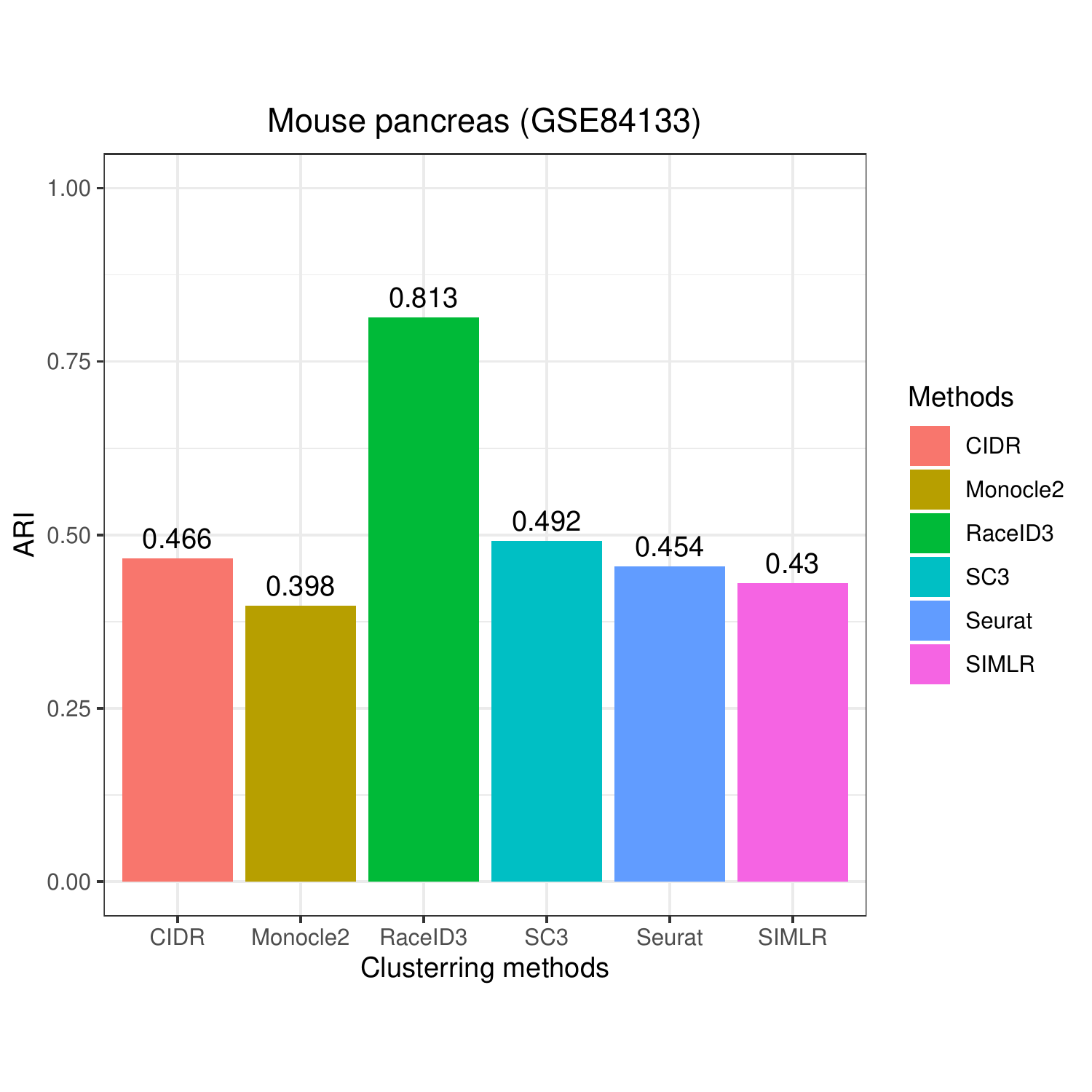}
	\includegraphics[width=0.5\textwidth]{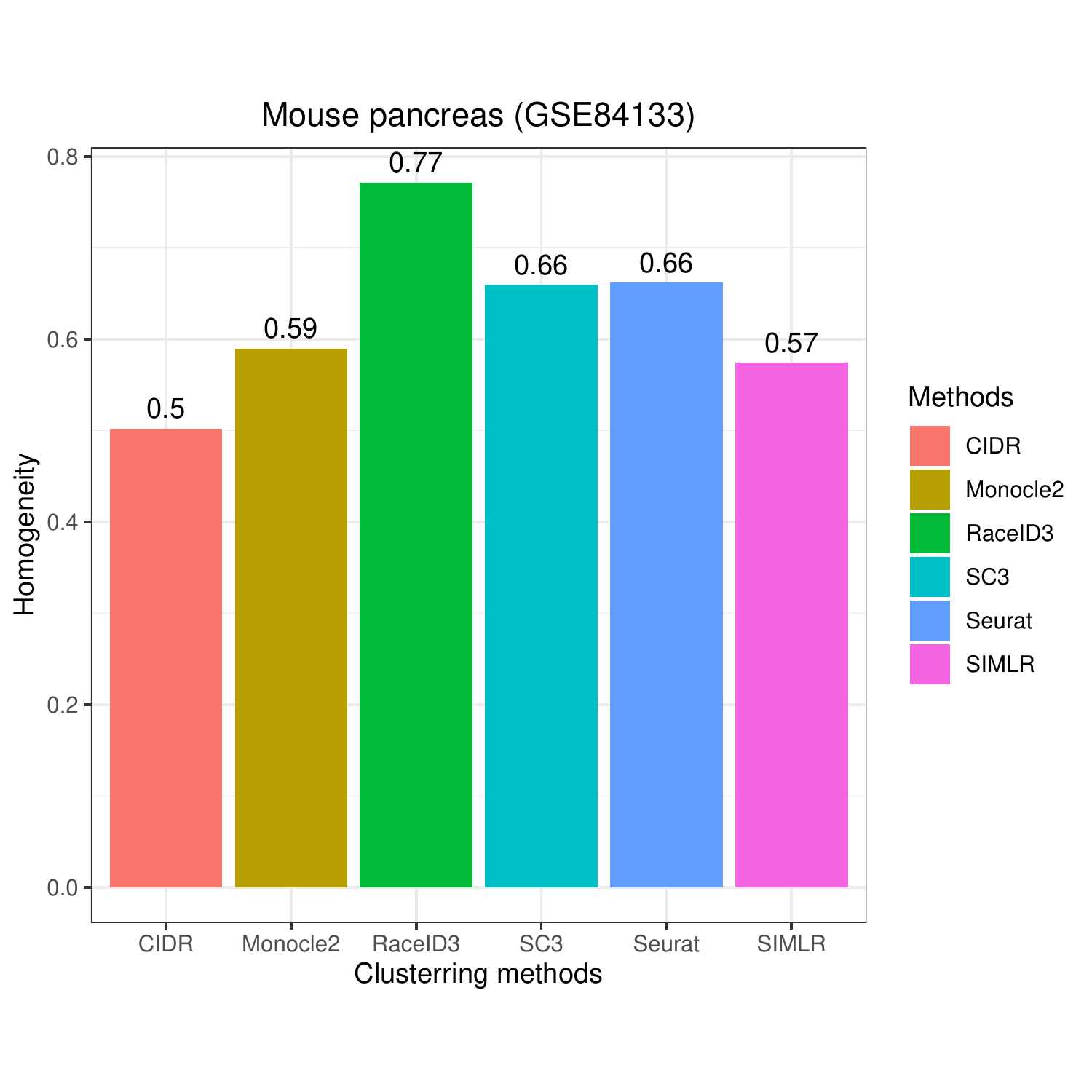}
    	\caption{Comparison of clustering performance on mouse pancreas single-cell RNA-seq data (GSE84133). The x-axis represents the clustering methods. The y-axis denotes the ARI or homogeneity scores of clustering results by RaceID3 \citep{Herman2018FateIDData}, Monocle2 \citep{Qiu2017ReversedTrajectories}, SIMLR \citep{Wang2017VisualizationLearning}, Seurat \citep{Satija2015SpatialData}, SC3 \citep{Kiselev2017SC3:Data}, and CIDR \citep{Lin2017CIDR:Data}.}
	\label{FIG:2}
\end{figure}

\begin{figure}
	\centering
	\includegraphics[width=0.5\textwidth]{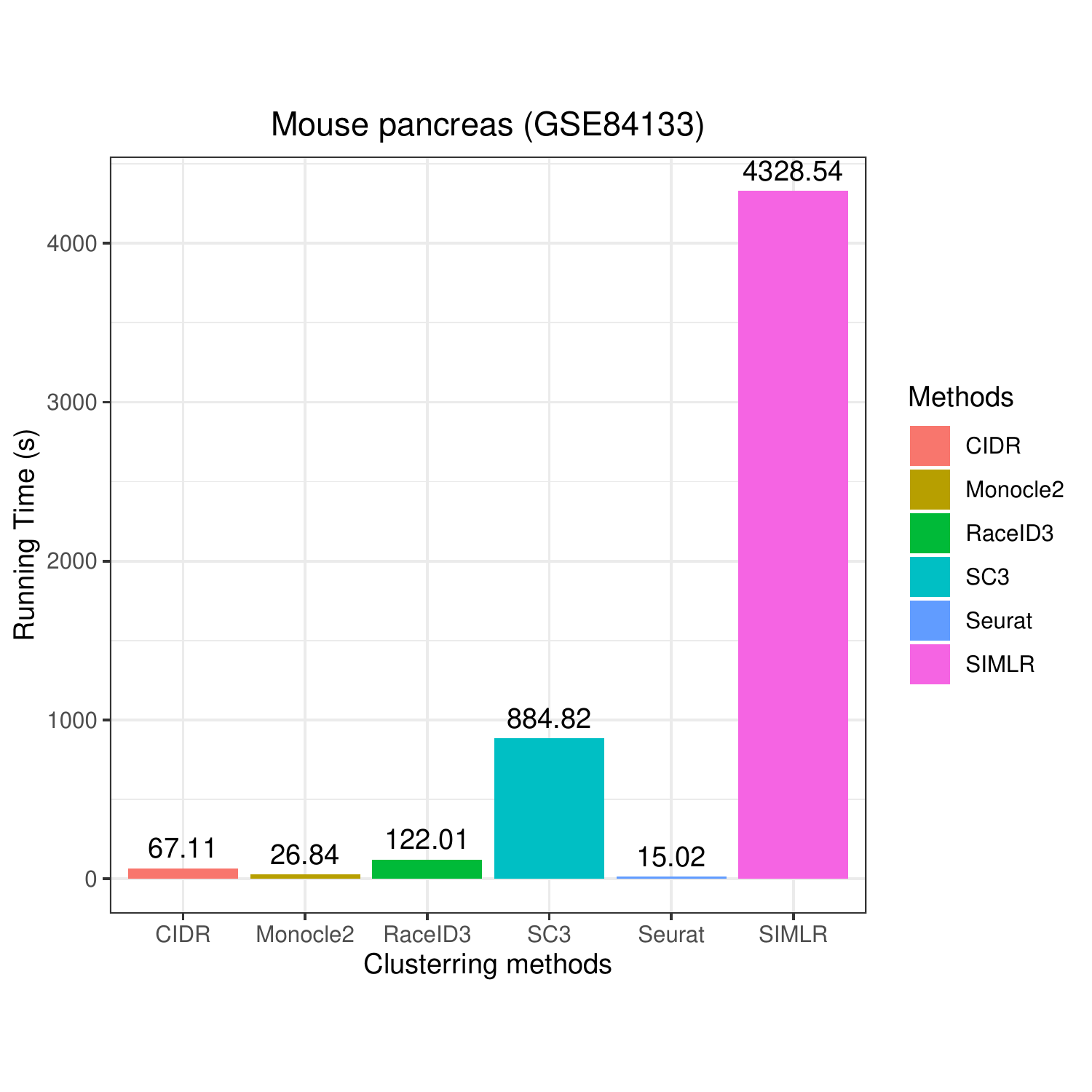}
    	\caption{Comparison of clustering efficiency on mouse pancreas single-cell RNA-seq data (GSE84133). The x-axis represents the clustering methods. The y-axis denotes the running time of clustering results by RaceID3 \citep{Herman2018FateIDData}, Monocle2 \citep{Qiu2017ReversedTrajectories}, SIMLR \citep{Wang2017VisualizationLearning}, Seurat \citep{Satija2015SpatialData}, SC3 \citep{Kiselev2017SC3:Data}, and CIDR \citep{Lin2017CIDR:Data}.}
	\label{FIG:3}
\end{figure}

\begin{figure*}
	\begin{minipage}[t]{0.5\textwidth}
		\centering
		\includegraphics[width=1\textwidth]{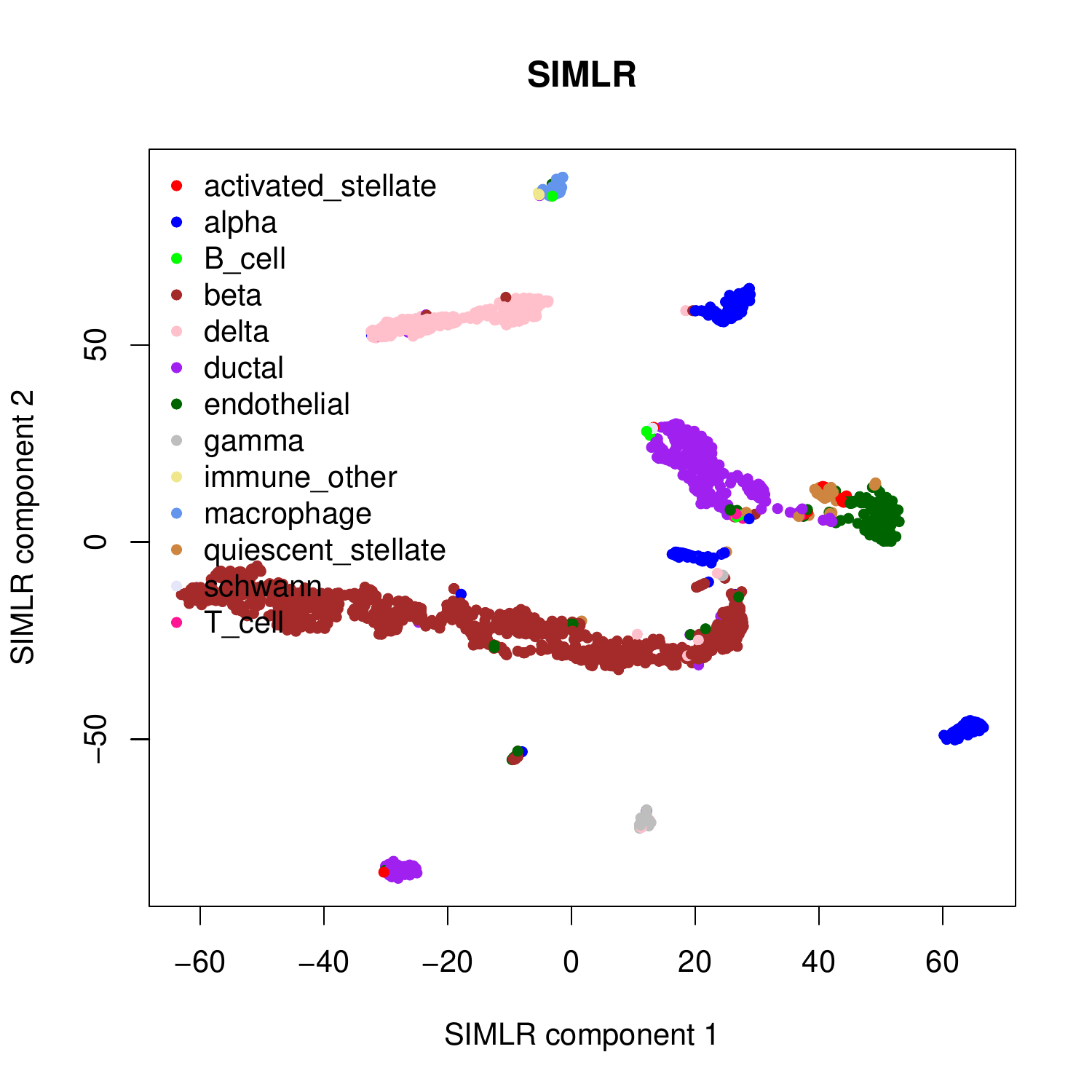}
		\label{fig:side:a}
	\end{minipage}%
	\begin{minipage}[t]{0.5\textwidth}
		\centering
		\includegraphics[width=1\textwidth]{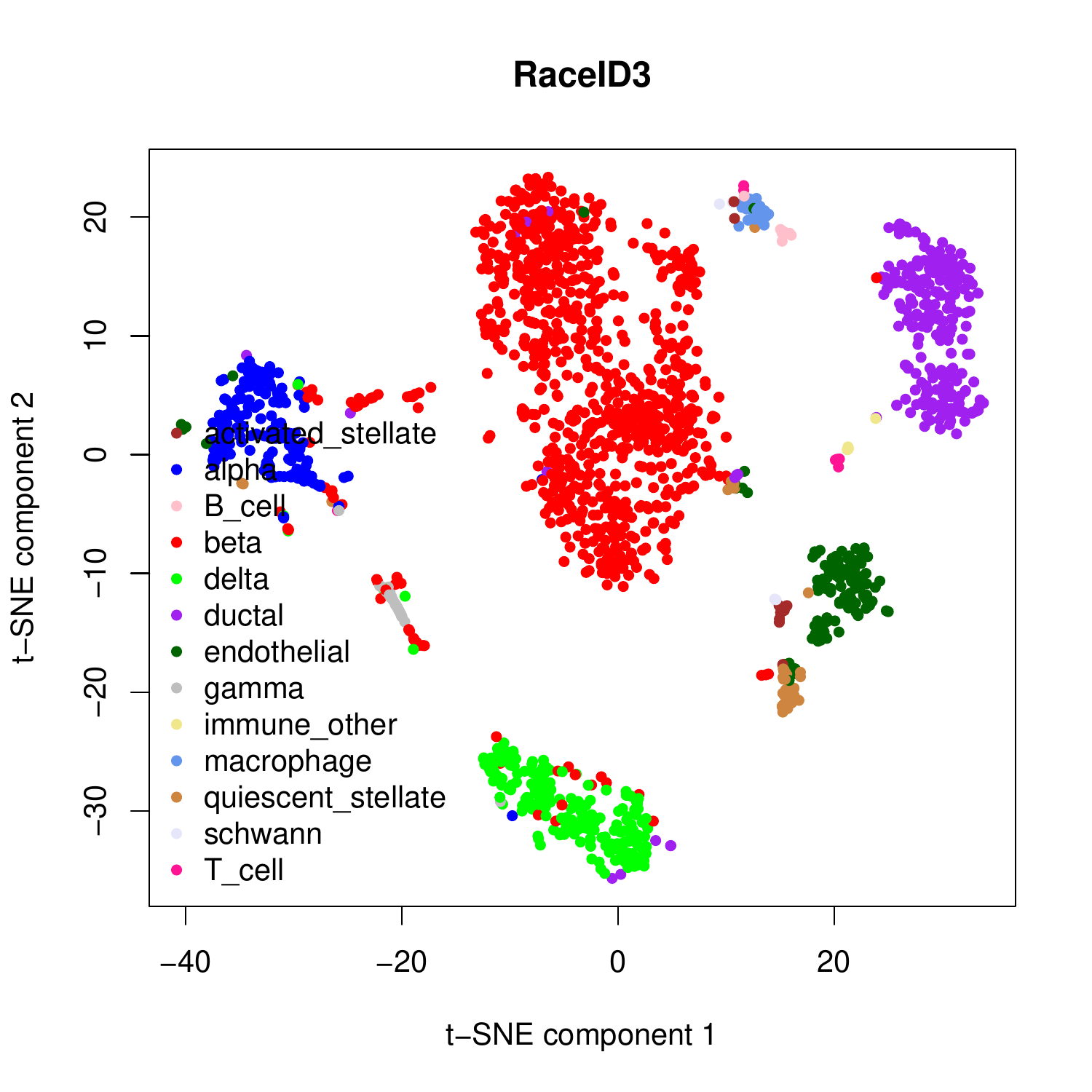}
		\label{fig:side:b}
	\end{minipage}\\
\end{figure*}
\addtocounter{figure}{-1}
\begin{figure*}
		\addtocounter{figure}{1}
		\begin{minipage}[t]{0.5\textwidth}
		\centering
		\includegraphics[width=1\textwidth]{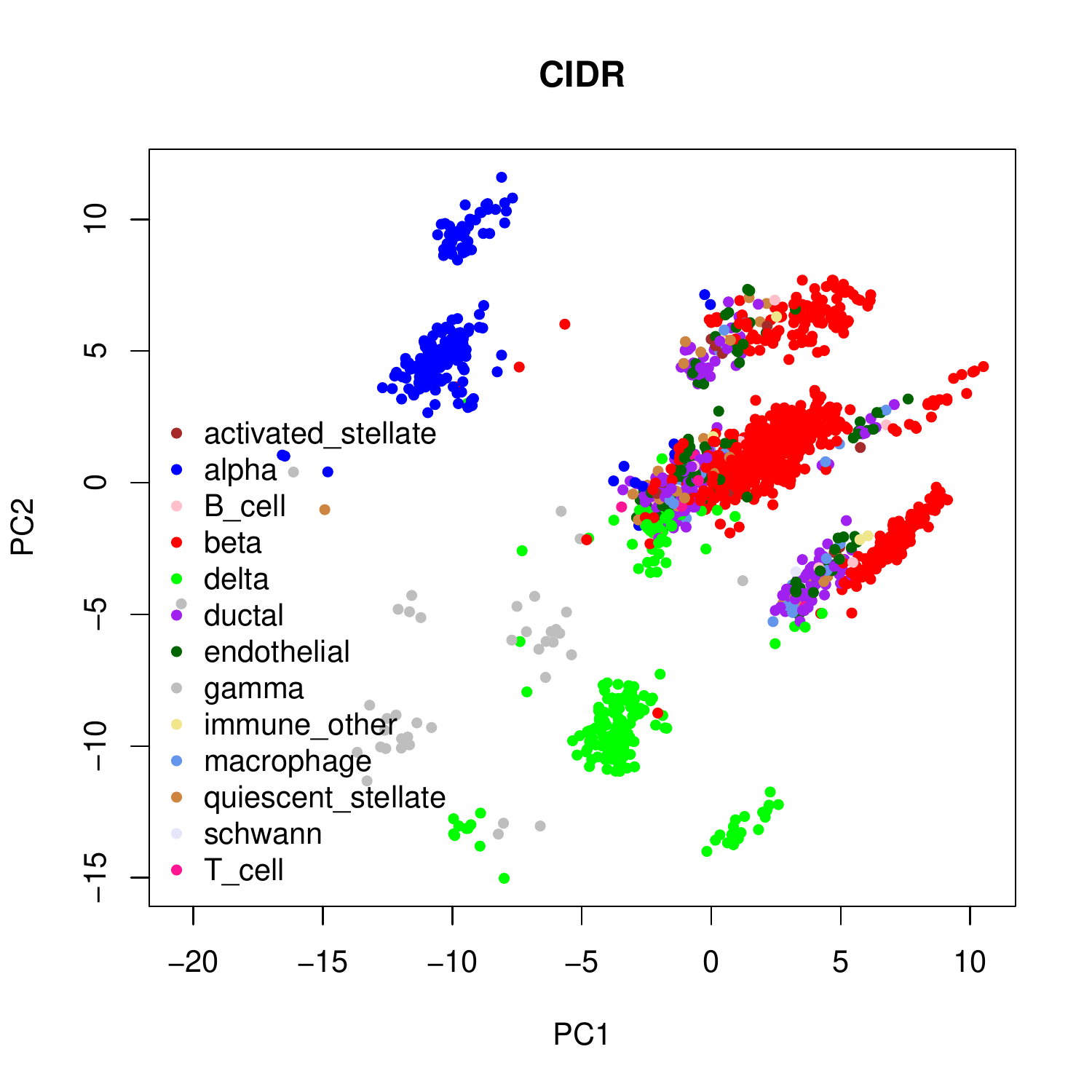}
		\label{fig:side:c}
	\end{minipage}%
	\begin{minipage}[t]{0.5\textwidth}
		\centering
		\includegraphics[width=1\textwidth]{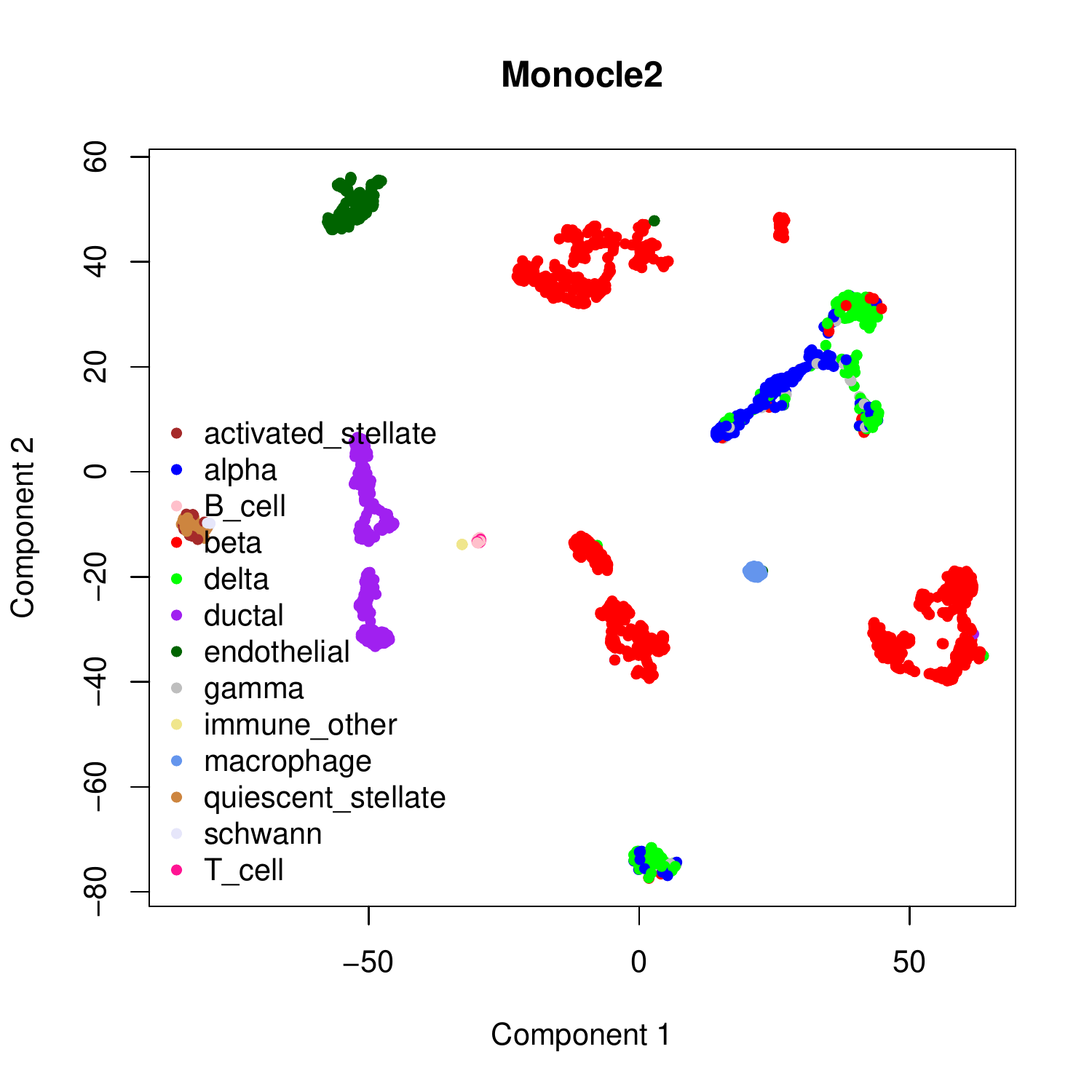}
		\label{fig:side:d}
	\end{minipage}\\
        \caption{Visualization of clustering performance on mouse pancreas single-cell RNA-seq data (GSE84133) from SIMLR \citep{Wang2017VisualizationLearning}, RaceID3 \citep{Herman2018FateIDData}, CIDR \citep{Lin2017CIDR:Data}, and Monocle2 \citep{Qiu2017ReversedTrajectories}.}	\label{FIG:4}
\end{figure*}

\begin{figure}
	\centering
	\includegraphics[width=0.5\textwidth]{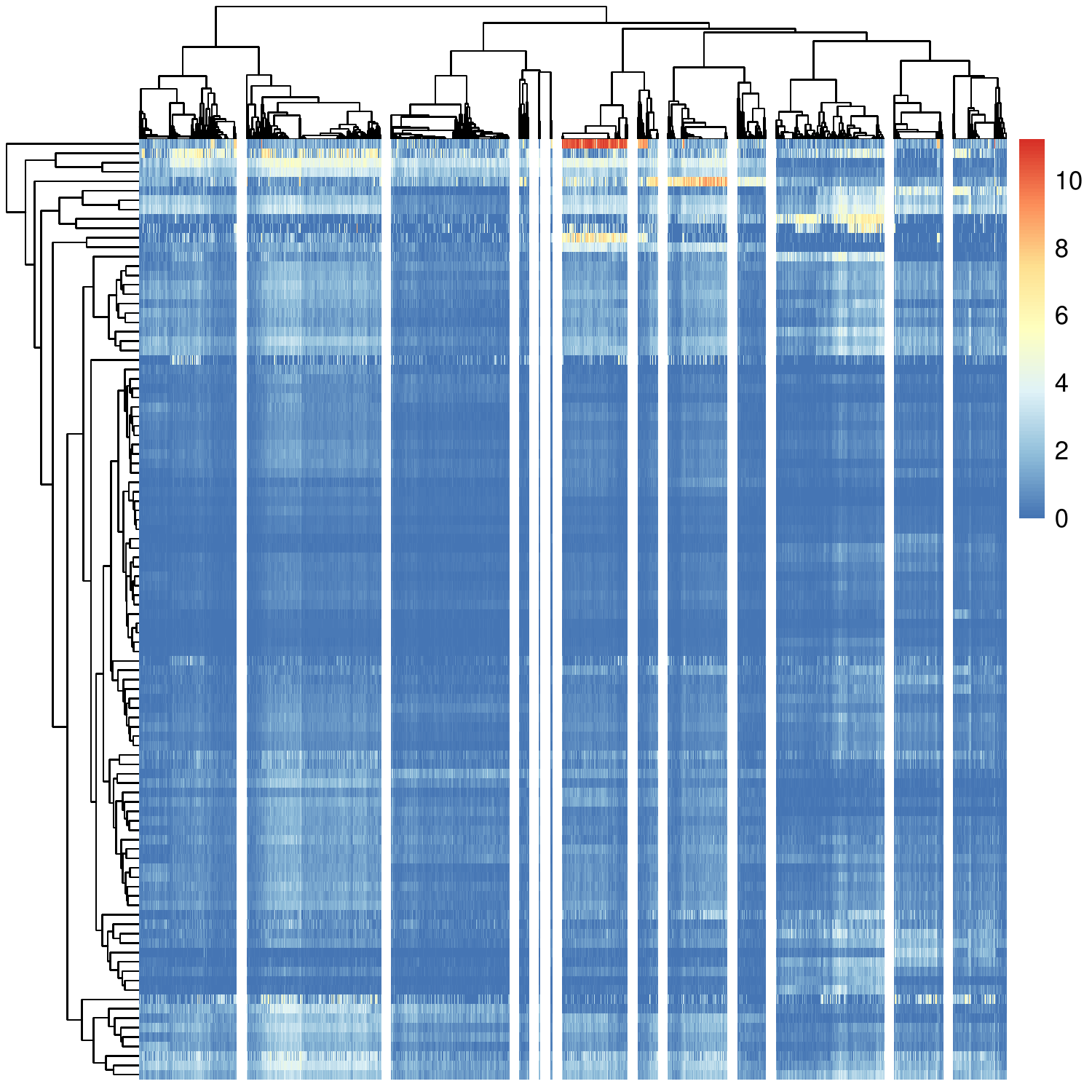}
	\caption{Visualization of clustering performance on mouse pancreas single-cell RNA-seq data (GSE84133) from SC3 \citep{Kiselev2017SC3:Data}. The x-axis represents cells. The y-axis denotes genes.}
	\label{FIG:5}
\end{figure}

\begin{figure}
	\centering
	\includegraphics[width=0.5\textwidth]{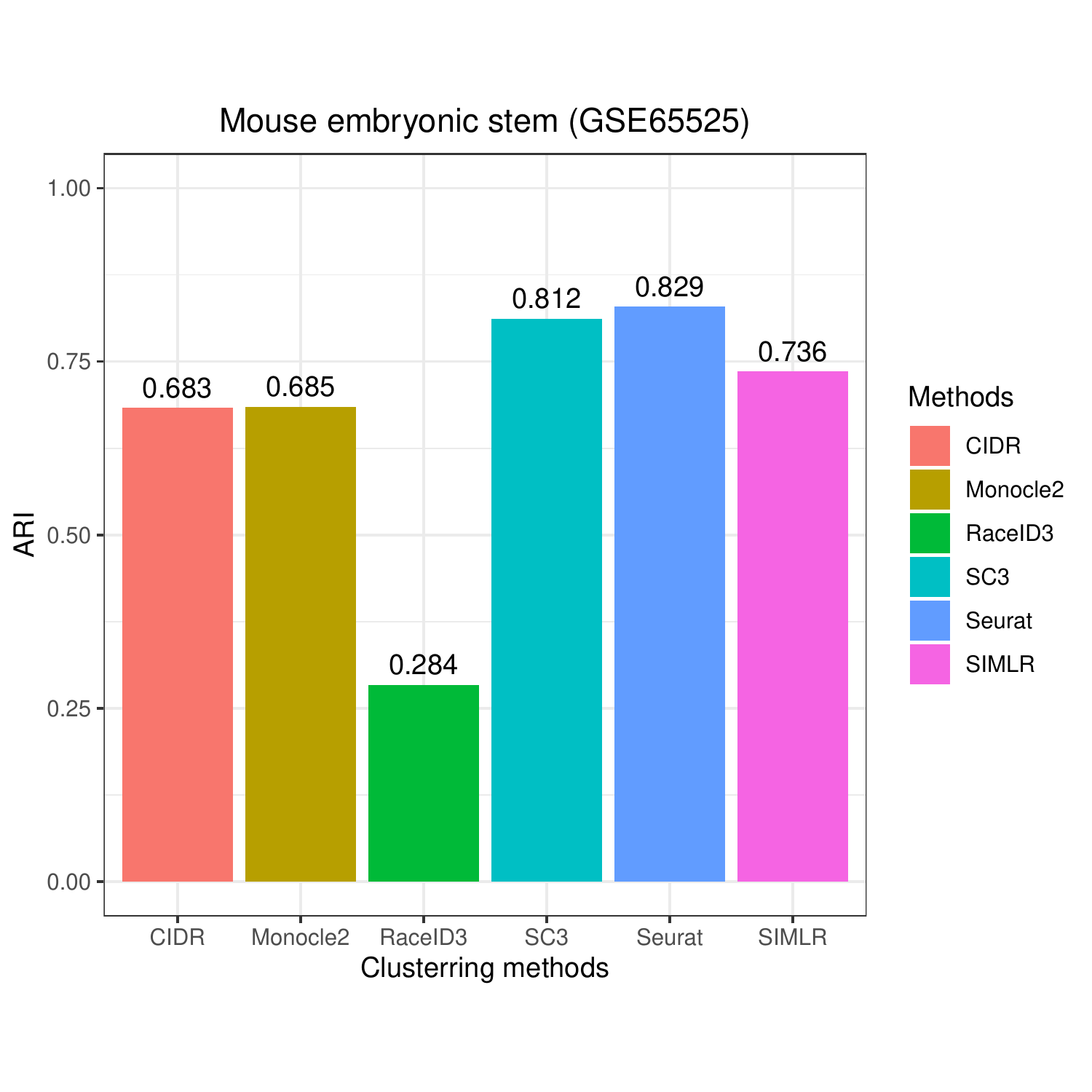}
	\includegraphics[width=0.5\textwidth]{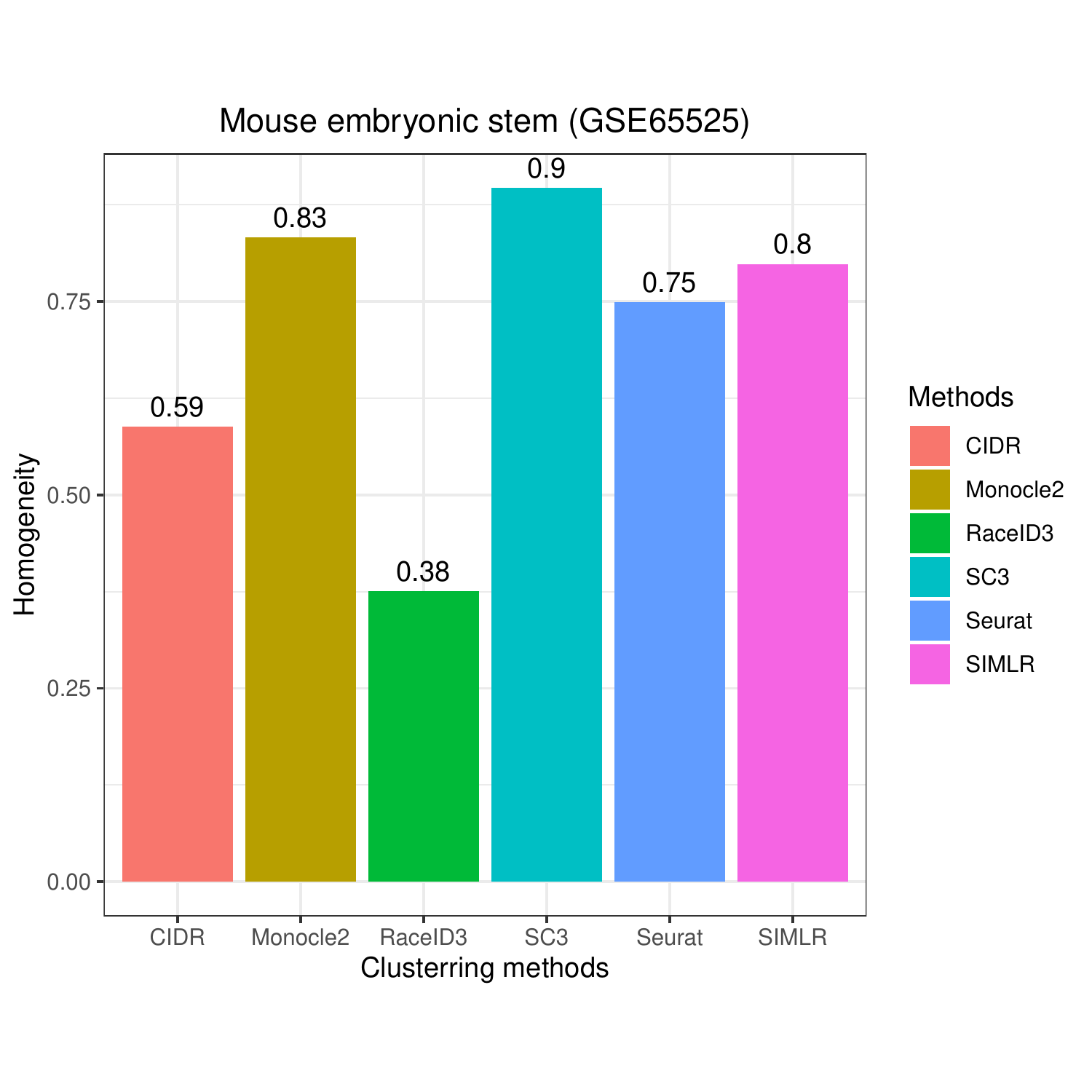}
	\caption{Comparison of clustering performance on mouse embryonic stem single-cell RNA-seq data (GSE65525). The x-axis represents the clustering methods. The y-axis denotes the ARI or homogeneity scores of clustering results across RaceID3 \citep{Herman2018FateIDData}, Monocle2 \citep{Qiu2017ReversedTrajectories}, SIMLR \citep{Wang2017VisualizationLearning}, Seurat \citep{Satija2015SpatialData}, SC3 \citep{Kiselev2017SC3:Data}, and CIDR \citep{Lin2017CIDR:Data}.}
	\label{FIG:6}
\end{figure}

\begin{figure}
	\centering
	\includegraphics[width=0.5\textwidth]{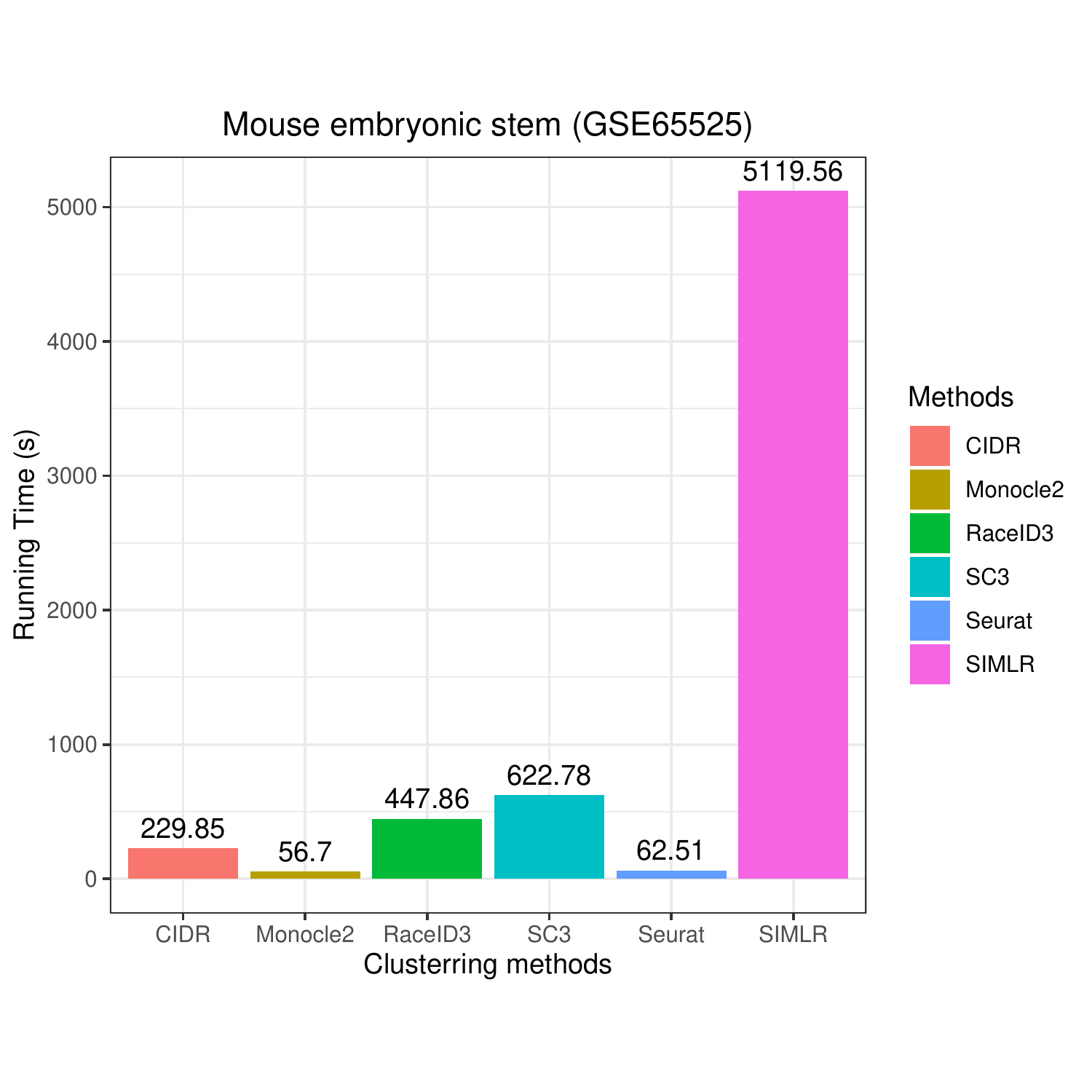}
	\caption{Comparison of clustering efficiency on mouse embryonic stem single-cell RNA-seq data (GSE65525). The x-axis represents the clustering methods. The y-axis denotes the running time of clustering results across RaceID3 \citep{Herman2018FateIDData}, Monocle2 \citep{Qiu2017ReversedTrajectories}, SIMLR \citep{Wang2017VisualizationLearning}, Seurat \citep{Satija2015SpatialData}, SC3 \citep{Kiselev2017SC3:Data}, and CIDR \citep{Lin2017CIDR:Data}.}
	\label{FIG:7}
\end{figure}

\begin{figure*}
	\begin{minipage}[t]{0.5\textwidth}
		\centering
		\includegraphics[width=1\textwidth]{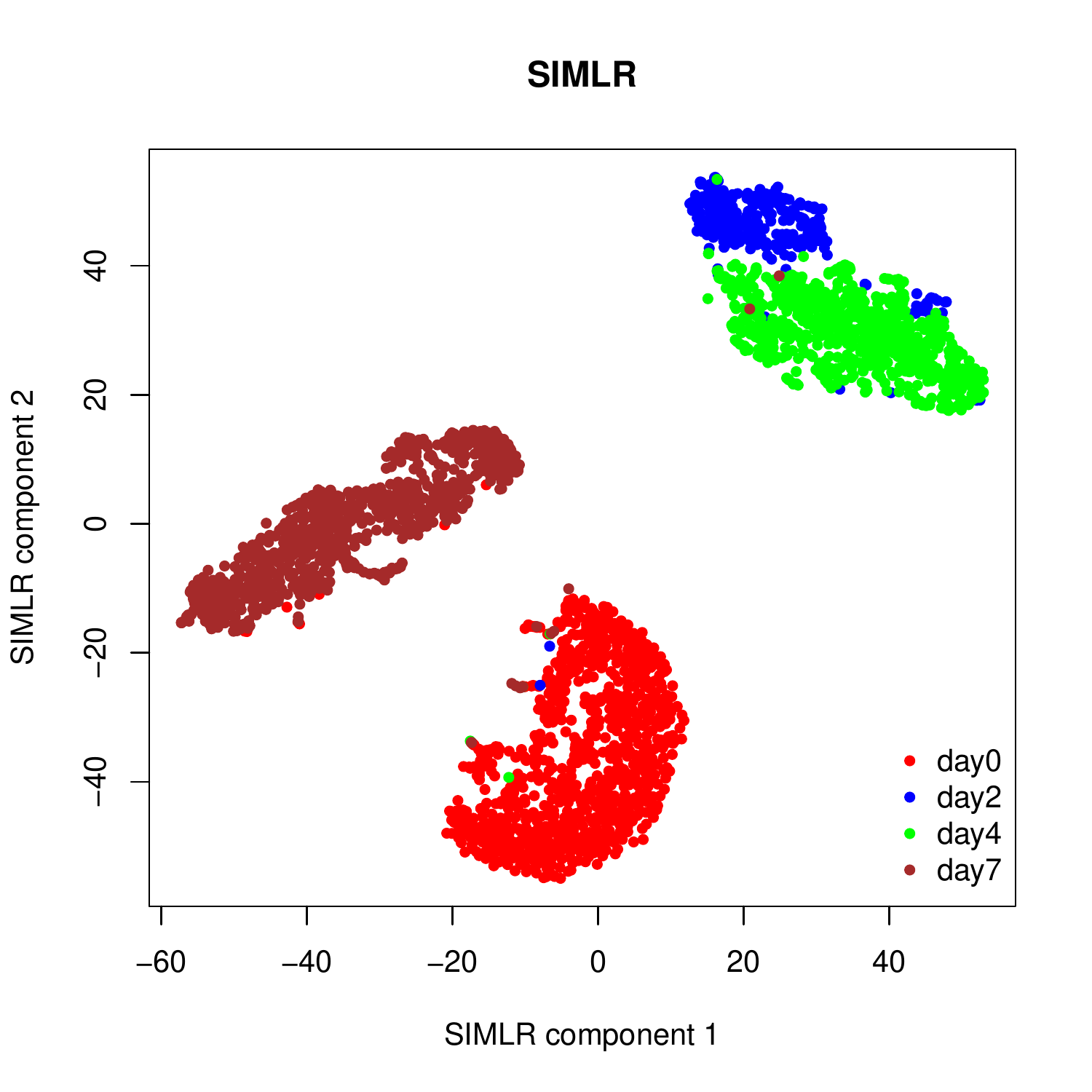}
		\label{fig:side:a}
	\end{minipage}%
	\begin{minipage}[t]{0.5\textwidth}
		\centering
		\includegraphics[width=1\textwidth]{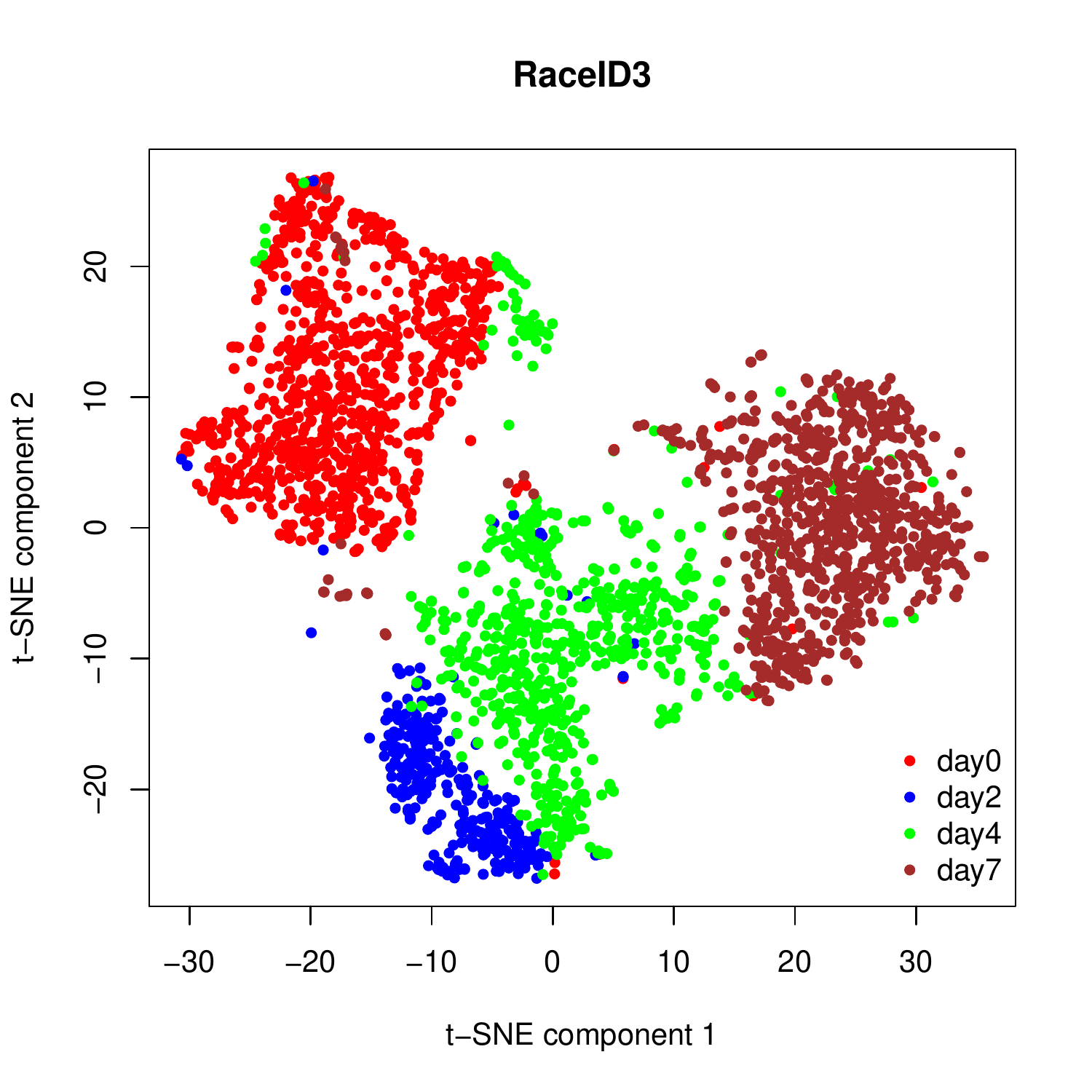}
		\label{fig:side:b}
	\end{minipage}\\
\end{figure*}
\addtocounter{figure}{-1}
\begin{figure*}
	\addtocounter{figure}{1}
	\begin{minipage}[t]{0.5\textwidth}
		\centering
		\includegraphics[width=1\textwidth]{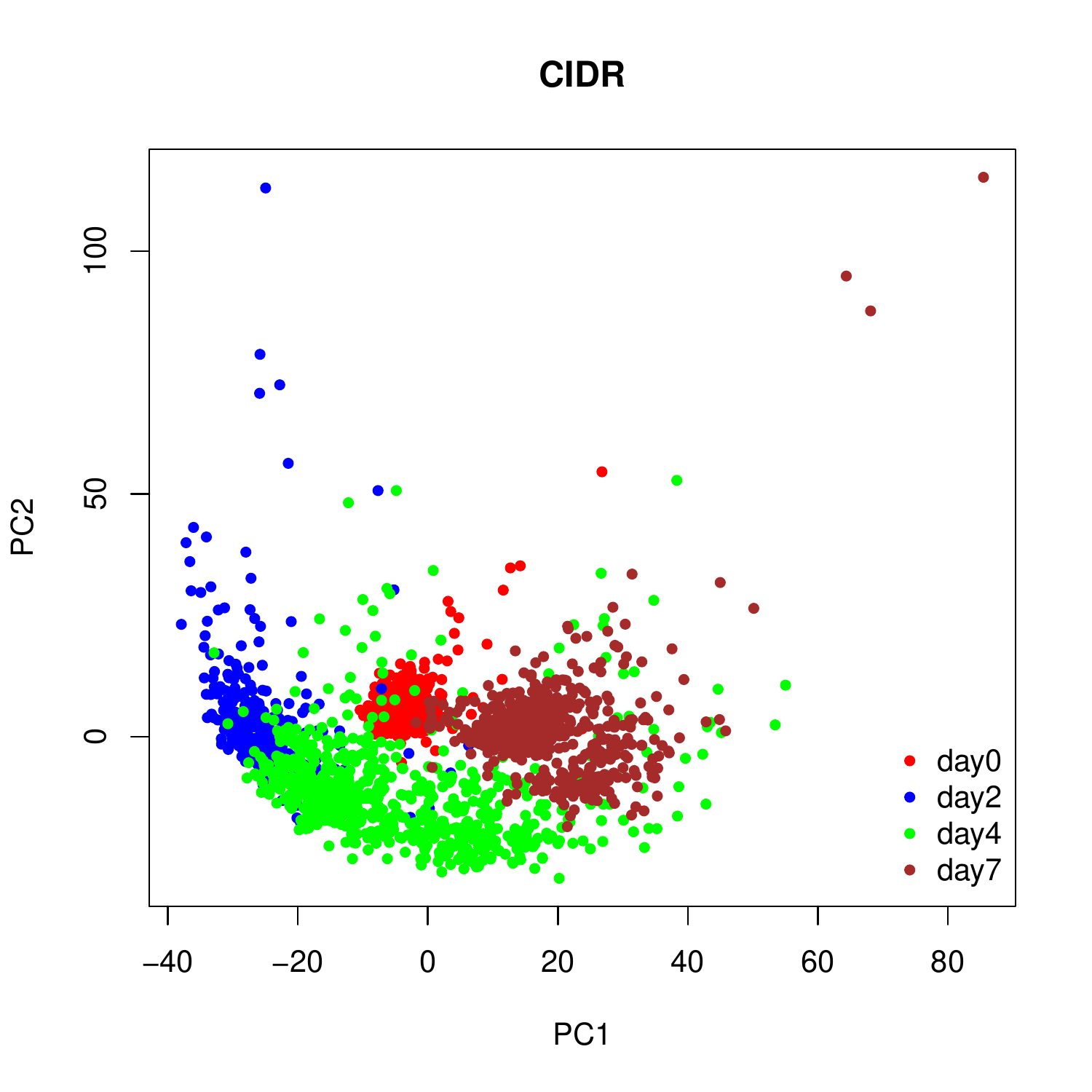}
		\label{fig:side:c}
	\end{minipage}%
	\begin{minipage}[t]{0.5\textwidth}
		\centering
		\includegraphics[width=1\textwidth]{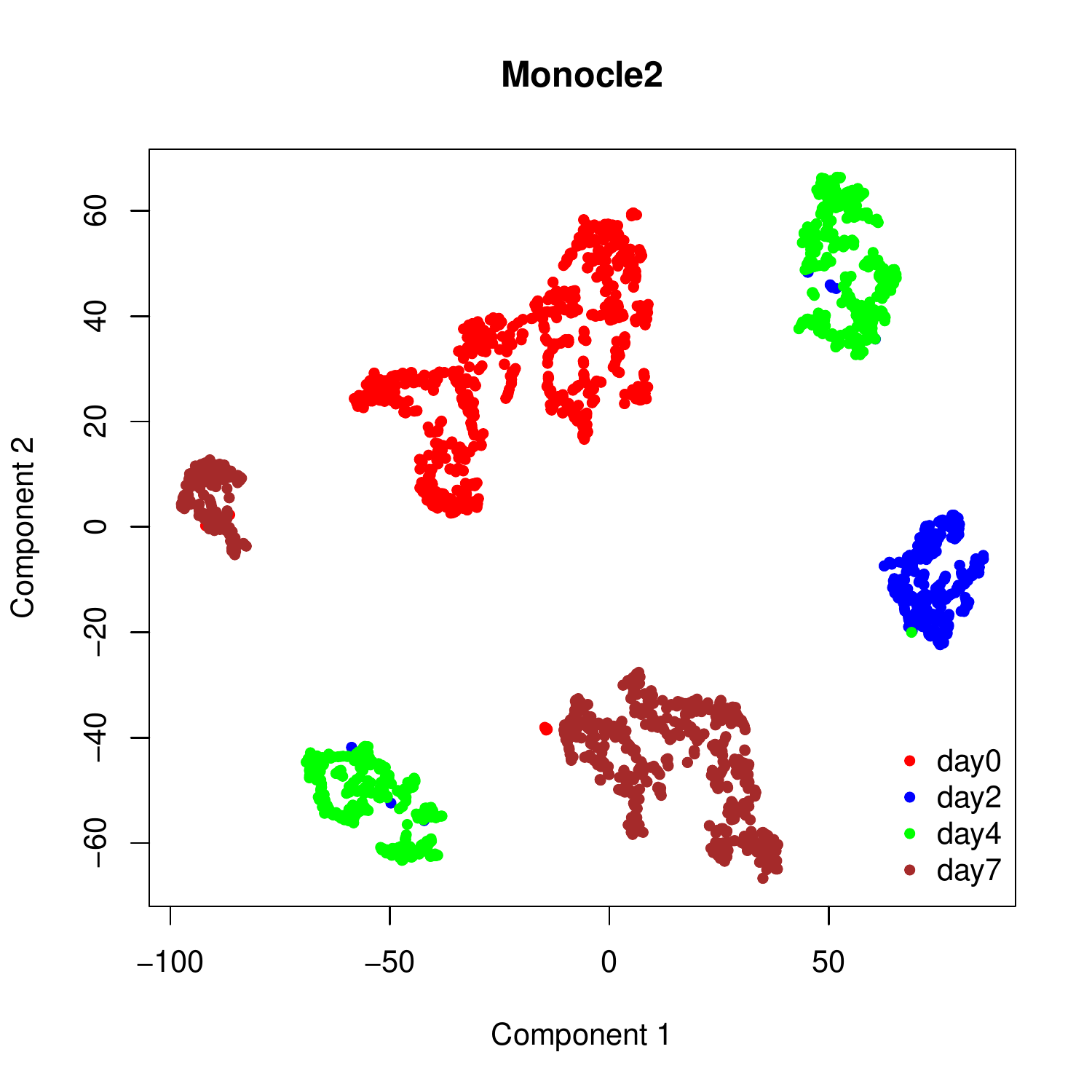}
		\label{fig:side:d}
	\end{minipage}\\
	\caption{Visualization of clustering performance on mouse embryonic stem single-cell RNA-seq large-scale dataset(GSE65525) by SIMLR \citep{Wang2017VisualizationLearning}, RaceID3 \citep{Herman2018FateIDData}, CIDR \citep{Lin2017CIDR:Data}, and Monocle2 \citep{Qiu2017ReversedTrajectories}.}	\label{FIG:8}
\end{figure*}

\begin{figure}
	\centering
	\includegraphics[width=0.5\textwidth]{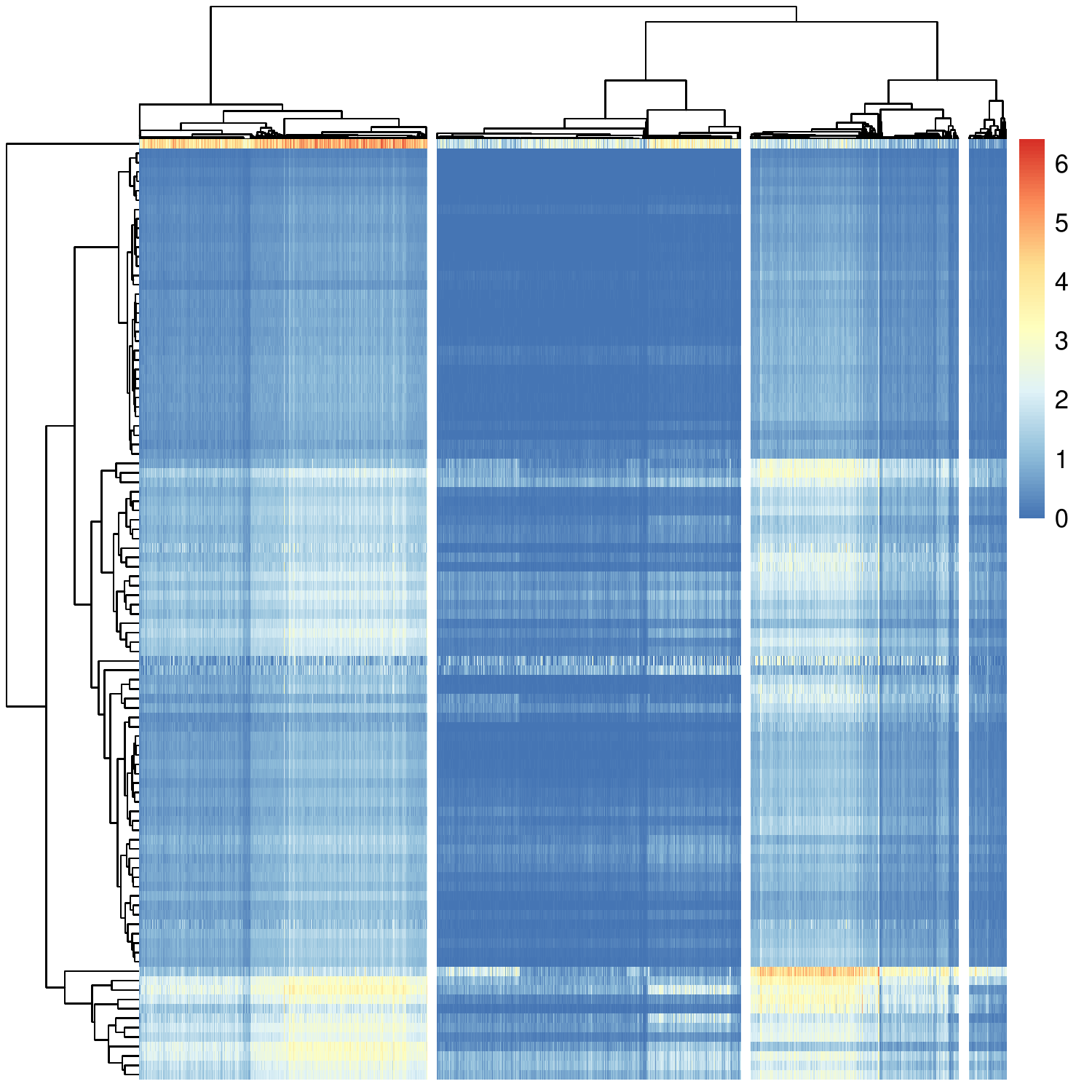}
	\caption{Visualization of clustering performance on mouse embryonic stem single-cell RNA-seq data (GSE65525) from SC3 \citep{Kiselev2017SC3:Data}. The x-axis represents cells. The y-axis denotes genes.}
	\label{FIG:9}
\end{figure}

In this section, we conduct independent experiments to evaluate several widely used single-cell RNA-seq clustering methods. Those clustering methods contain RaceID3 \citep{Herman2018FateIDData}, Monocle2 \citep{Qiu2017ReversedTrajectories}, SIMLR \citep{Wang2017VisualizationLearning}, Seurat \citep{Satija2015SpatialData}, SC3 \citep{Kiselev2017SC3:Data}, and CIDR \citep{Lin2017CIDR:Data}. We applied six single-cell RNA-seq clustering methods on two different droplet-based transcriptomic datasets (GSE84133 and GSE65525) with cell types annotations. For the evaluation and comparison, we introduce two commonly used metrics including Adjusted Rand index, Running Time, and Homogeneity Score to measure the clustering performance and efficiency respectively. The parameter setting of the cluster methods on both datasets are tabulated in Table~\ref{tbl2}. In particular, we would like to note that most of parameters were chosen based on the default setting given by individual methods.

\subsection{Evaluation metrics for clustering}
Since the single-cell RNA-seq clustering is an unsupervised learning task in most studies, three common metrics Adjusted Rand index, Running Time, and Homogeneity Score are introduced for the evaluation. 

Adjusted Rand index (ARI) proposed by \citet{Hubert1985ComparingPartitions} can be used to measure the similarity between the clustering results of interest and the true clustering. However, ARI is widely applied as the metric of single-cell RNA-seq clustering only when the cell-labels are available \citep{Kiselev2017SC3:Data,Ntranos2016FastCounts,Lin2017CIDR:Data,Aibar2017SCENIC:Clustering,Xu2015IdentificationMethod}. Given a set of $n$ cells and two clusterings ($X = \{X_1, X_2, ..., X_s\}$ partitioned by clustering method and $Y = \{Y_1, Y_2, ..., Y_r\}$ partitioned by annotated cell types) of these cells, the overlap between the two clusterings can be summarized in a contingency table with $s$ rows and $r$ columns. The ARI is defined as below.

\begin{equation} \label{Eq:01}
ARI=\frac{\sum_{ij}\binom{n_{ij}}{2}-[\sum_{i}\binom{a_i}{2}\sum_{j}\binom{b_j}{2}]/\binom{n}{2}}{\frac{1}{2}[\sum_{i}\binom{a_i}{2}+\sum_{j}\binom{b_j}{2}]-[\sum_{i}\binom{a_i}{2}\sum_{j}\binom{b_j}{2}]/\binom{n}{2}}
\end{equation}
where $n_{ij} = |X_i \cap Y_j|$ denotes the values from the contingency table; $a_i = \sum_{j}n_{ij}$ and $b_j = \sum_{i}n_{ij}$ represent the $i$th row sums and $j$th column sums of the contingency table, respectively; and $\binom{}{}$ denotes a binomial coefficient. ARI = 1 indicates a perfect overlap between clusters $X$ and $Y$, while ARI = 0 indicates random clustering. 

Homogeneity Score \citep{Rosenberg2007V-Measure:Measure} evaluates the performance of clustering results with regards to the ground truth. It is defined as:

\begin{equation} \label{Eq:02}
Homogeneity=\frac{I(X,Y)}{H(Y)}
\end{equation}
where $H(Y)=I(Y,Y)$ is the entropy of $Y$ and $I(X,Y)$ is the mutual information of $X$ and $Y$. It is bounded between 0 and 1. Homogeneity = 1 indicates all of its clusters contain only data points from a single class while low values indicate that clusters contain mixed known groups.

In addition, running time is usually measured to evaluate the algorithm efficiency. High efficiency is an important feature since the single-cell RNA-seq data usually come up with thousands of cells and genes. 

\subsection{Performance on mouse pancreas single-cell RNA-seq dataset (GSE84133)}

In mouse pancreas single-cell RNA-seq dataset (GSE84133) \citep{Baron2016AStructure}, there are 1,886 cells in 13 cell types after the exclusion of hybrid cells. GSE84133 has 14,878 genes. Figure~\ref{FIG:2} and \ref{FIG:3} shows the ARI, homogeneity scores, and running time of RaceID3, Monocle2, SIMLR, Seurat, SC3, and CIDR on GSE84133 for performance comparision. The results show that RacID3 exhibit the best ARI (=0.813) and homogeneity (=0.77) performance among the six methods. The ARI of other methods does not exceed 0.500. SIMLR is a time-consuming method and it took 1.20 hours to conduct the clustering task. However, CIDR can only identify seven cell types from GSE84133 since it belongs to hierarchical clustering and is unable to predetermine the number of clusters. SC3, SIMLR, and Monocle2 cannot provide an accurate estimation of the cluster count and it has to be determined manually. Seurat, Monocle2, and RaceID3 require user to adjust multiple parameters to achieve the best clustering performance that affected the user friendliness. Figure~\ref{FIG:4} illustrates the clustering performance of SIMLR, RaceID3, CIDR, and Monocle2 on GSE84133. The visualization results are directly obtained from their R packages. Figure~\ref{FIG:5} displays the clustering results from SC3. Since SC3 belongs to hierarchical clustering, the clustering result is illustrated in heatmap and it is set to show the 13 cell types.

\subsection{Performance on mouse embryonic stem single-cell RNA-seq dataset (GSE65525)}

In mouse embryonic stem single-cell RNA-seq large-scale dataset (GSE65525) \citep{Klein2015DropletCells}, there are 2717 cells in four annotated cell types. GSE65525 has 24,175 genes. Figure~\ref{FIG:6} and \ref{FIG:7} depict the ARI, homogeneity scores, and running time of RaceID3, Monocle2, SIMLR, Seurat, SC3, and CIDR on GSE65525 for performance comparison. The results show that Seurat exhibits the best ARI (=0.829) performance among the six methods, while its homogeneity score (homogeneity=0.75) is lower than SC3 (homogeneity=0.9). The ARI of SC3 (0.812) also exceeds 0.80 and it achieve higher homogeneity than others. Hence, SC3 exhibits robust clustering performance on large datasets with a reasonable sacrifice on efficiency. The ARI and homogeneity scores of Monocle2, SIMLR, RaceID3, SC3, and CIDR showed different degrees of accuracy. RaceID achieved the worst ARI (=0.284) and homogeneity score (homogeneity=0.38) across six methods on GSE65525. SIMLR has been run for 85 minutes which took far more than other five methods. Hence, the results show that RaceID3 may not be suitable to large-scale single-cell RNA-seq datasets. Figure~\ref{FIG:8} illustrates the clustering performance of SIMLR, RaceID, CIDR, and Monocle2 on GSE65525. The visualization results are directly obtained from their R packages. Results from Figure~\ref{FIG:8} show that all methods result in different degrees of undesirable overlaps between clusters. Figure~\ref{FIG:9} displays the clustering results of SC3 on GSE65525 and it shows the four correct cell types.

\begin{table*}[width=1.8\linewidth, pos=h]
	\caption{Parameter setting of clustering methods on GSE65525 (GSE84133).}\label{tbl2}
	\begin{tabular*}{\tblwidth}{@{}LL@{} }
		\toprule
		Methods (Version)    & Limitations\\
		\midrule
		\tabincell{L}{RaceID3 (0.1.6)\\ \citep{Herman2018FateIDData}}       & \tabincell{L}{mintotal = 100, minexpr = 5, minnumber = 1, maxexpr = 500, \\downsample = FALSE dsn = 1, rseed = 17000, clustnr = 20, bootnr = 50 \\
	     metric = "pearson", do.gap = TRUE, SE.factor = .25, B.gap = 50, \\cln=0, rseed = 18000} \\\hline
		\tabincell{L}{SC3 (1.14.0)\\ \citep{Kiselev2017SC3:Data}}            & \tabincell{L}{ks = 4 (13), biology = FALSE, rand\_seed = sample(1:10000,1)} \\\hline
		\tabincell{L}{CIDR (0.1.5)\\ \citep{Lin2017CIDR:Data}}              &  \tabincell{L}{min1 = 3, min2 = 8,
        N = 2000, alpha = 0.1, fast = TRUE} \\\hline
		\tabincell{L}{SIMLR (1.12.0)\\ \citep{Wang2017VisualizationLearning}}& \tabincell{L}{c = 4 (13), normalize = TRUE, k = 10, if.impute = FALSE, cores.ratio = 1} \\\hline
		\tabincell{L}{Seurat (3.1.1)\\ \citep{Satija2015SpatialData}}       & \tabincell{L}{normalization.method = "LogNormalize", scale.factor = 100, npcs = 100\\ vars.to.regress = "percent.mt", ndims.print = 1:5, nfeatures.print = 5\\reduction = "pca", dims = 1:75, nn.eps = 0.5, resolution = 1} \\\hline
		\tabincell{L}{Monocle2 (2.14.0)\\ \citep{Qiu2017ReversedTrajectories}} & \tabincell{L}{max\_components = 3, num\_dim = 10, reduction\_method = 'tSNE',\\ verbose = T, check\_duplicates = F, num\_clusters = 4 (13)}  \\
		\bottomrule
	\end{tabular*}
\end{table*}

\section{Discussions and conclusions}

Single-cell RNA-seq data analysis is a crucial component in whole-transcriptome studies. In particular, data clustering is the central component of single-cell RNA-seq analysis. Clustering results can affect the performance of downstream analysis including identifying rare or new cell types, gene expression patterns that are predictive of cellular states, and functional implications of stochastic transcription. There are several related studies for the performance evaluation of clustering methods on single-cell RNA-seq data \citep{Duo2018AData,Freytag2018ComparisonData}. Those studies focused on assessing the methods for clustering single-cell RNA-seq data, while the data preprocessing steps may not be included in the respective discussion section, although it could have significantly influences on the downstream clustering performance. Therefore, in this study, we reviewed several clustering methods. In addition, the upstream RNA-seq data preprocessing steps have also been reviewed since those steps can significantly affect the downstream clustering performance. Lastly, our performance comparison experiments have also been conducted, revealing independent insights into the state-of-the-arts methods without any conflict of interest. 

Those clustering methods show expected performance on single-cell RNA-seq data. However, those clustering methods have its drawbacks; for instance, {\it k}-means clustering require users to determine the number of clusters and is sensitive to outliers; hierarchical clustering has high complexity and may be unsuitable to large-scale single-cell RNA-seq data; community-detection-based clustering cannot provide the estimation of number of clusters and is unsuitable for small communities; density-based clustering has advantages in detecting rare cell types with a sacrifice on large cluster performance. 

In addition to those limitations, there are still some technical challenges in single-cell RNA-seq clustering. With the advanced development of single-cell RNA-seq techniques, the single-cell datasets are growing to be extremely high-dimensional and sparse. Although some methods can deal with those data in a time span of hours such as SIMLR, visualization of those data is still a challenge.   Moreover, the low dimensionality of expression profiles implies intensive gene co-expression signature that may inspire us to develop new clustering methods on low-dimensional data to interpret cell types \citep{Crow2018Co-expressionSin}. Advanced data integration and analysis approaches are needed for both basic research and clinical studies in the coming years.

\subsection*{Ethics approval and consent to participate}
Not applicable.

\subsection*{Conflict of Interest}
The authors declare no competing interests.


\bibliographystyle{cas-model2-names}

\bibliography{cas-refs}

\begin{thebibliography}{69}
\expandafter\ifx\csname natexlab\endcsname\relax\def\natexlab#1{#1}\fi
\providecommand{\url}[1]{\texttt{#1}}
\providecommand{\href}[2]{#2}
\providecommand{\path}[1]{#1}
\providecommand{\DOIprefix}{doi:}
\providecommand{\ArXivprefix}{arXiv:}
\providecommand{\URLprefix}{URL: }
\providecommand{\Pubmedprefix}{pmid:}
\providecommand{\doi}[1]{\href{http://dx.doi.org/#1}{\path{#1}}}
\providecommand{\Pubmed}[1]{\href{pmid:#1}{\path{#1}}}
\providecommand{\bibinfo}[2]{#2}
\ifx\xfnm\relax \def\xfnm[#1]{\unskip,\space#1}\fi
\bibitem[{Aibar et~al.(2017)Aibar, Gonz{\'{a}}lez-Blas, Moerman, Huynh-Thu,
  Imrichova, Hulselmans, Rambow, Marine, Geurts, Aerts, van~den Oord, Atak,
  Wouters and Aerts}]{Aibar2017SCENIC:Clustering}
\bibinfo{author}{Aibar, S.}, \bibinfo{author}{Gonz{\'{a}}lez-Blas, C.B.},
  \bibinfo{author}{Moerman, T.}, \bibinfo{author}{Huynh-Thu, V.A.},
  \bibinfo{author}{Imrichova, H.}, \bibinfo{author}{Hulselmans, G.},
  \bibinfo{author}{Rambow, F.}, \bibinfo{author}{Marine, J.C.},
  \bibinfo{author}{Geurts, P.}, \bibinfo{author}{Aerts, J.},
  \bibinfo{author}{van~den Oord, J.}, \bibinfo{author}{Atak, Z.K.},
  \bibinfo{author}{Wouters, J.}, \bibinfo{author}{Aerts, S.},
  \bibinfo{year}{2017}.
\newblock \bibinfo{title}{{SCENIC: single-cell regulatory network inference and
  clustering}}.
\newblock \bibinfo{journal}{Nature Methods} \bibinfo{volume}{14},
  \bibinfo{pages}{1083--1086}.
\bibitem[{Anders and Huber(2010)}]{Anders2010DifferentialData}
\bibinfo{author}{Anders, S.}, \bibinfo{author}{Huber, W.},
  \bibinfo{year}{2010}.
\newblock \bibinfo{title}{{Differential expression analysis for sequence count
  data}}.
\newblock \bibinfo{journal}{Genome Biology} \bibinfo{volume}{11},
  \bibinfo{pages}{R106}.
\bibitem[{Andrews and Hemberg(2018)}]{Andrews2018IdentifyingScRNASeq}
\bibinfo{author}{Andrews, T.S.}, \bibinfo{author}{Hemberg, M.},
  \bibinfo{year}{2018}.
\newblock \bibinfo{title}{{Identifying cell populations with scRNASeq}}.
\newblock \bibinfo{journal}{Molecular Aspects of Medicine}
  \bibinfo{volume}{59}, \bibinfo{pages}{114--122}.
\bibitem[{Bacher et~al.(2017)Bacher, Chu, Leng, Gasch, Thomson, Stewart, Newton
  and Kendziorski}]{Bacher2017SCnorm:Data}
\bibinfo{author}{Bacher, R.}, \bibinfo{author}{Chu, L.F.},
  \bibinfo{author}{Leng, N.}, \bibinfo{author}{Gasch, A.P.},
  \bibinfo{author}{Thomson, J.A.}, \bibinfo{author}{Stewart, R.M.},
  \bibinfo{author}{Newton, M.}, \bibinfo{author}{Kendziorski, C.},
  \bibinfo{year}{2017}.
\newblock \bibinfo{title}{{SCnorm: robust normalization of single-cell RNA-seq
  data}}.
\newblock \bibinfo{journal}{Nature Methods} \bibinfo{volume}{14},
  \bibinfo{pages}{584--586}.
\bibitem[{Baron et~al.(2016)Baron, Veres, Wolock, Faust, Gaujoux, Vetere, Ryu,
  Wagner, Shen-Orr, Klein, Melton and Yanai}]{Baron2016AStructure}
\bibinfo{author}{Baron, M.}, \bibinfo{author}{Veres, A.},
  \bibinfo{author}{Wolock, S.L.}, \bibinfo{author}{Faust, A.L.},
  \bibinfo{author}{Gaujoux, R.}, \bibinfo{author}{Vetere, A.},
  \bibinfo{author}{Ryu, J.H.}, \bibinfo{author}{Wagner, B.K.},
  \bibinfo{author}{Shen-Orr, S.S.}, \bibinfo{author}{Klein, A.M.},
  \bibinfo{author}{Melton, D.A.}, \bibinfo{author}{Yanai, I.},
  \bibinfo{year}{2016}.
\newblock \bibinfo{title}{{A Single-Cell Transcriptomic Map of the Human and
  Mouse Pancreas Reveals Inter- and Intra-cell Population Structure}}.
\newblock \bibinfo{journal}{Cell Systems} \bibinfo{volume}{3},
  \bibinfo{pages}{346--360}.
\bibitem[{Becht et~al.(2019)Becht, McInnes, Healy, Dutertre, Kwok, Ng, Ginhoux
  and Newell}]{Becht2019DimensionalityUMAP}
\bibinfo{author}{Becht, E.}, \bibinfo{author}{McInnes, L.},
  \bibinfo{author}{Healy, J.}, \bibinfo{author}{Dutertre, C.A.},
  \bibinfo{author}{Kwok, I.W.}, \bibinfo{author}{Ng, L.G.},
  \bibinfo{author}{Ginhoux, F.}, \bibinfo{author}{Newell, E.W.},
  \bibinfo{year}{2019}.
\newblock \bibinfo{title}{{Dimensionality reduction for visualizing single-cell
  data using UMAP}}.
\newblock \bibinfo{journal}{Nature Biotechnology} \bibinfo{volume}{37},
  \bibinfo{pages}{38--47}.
\bibitem[{Blondel et~al.(2008)Blondel, Guillaume, Lambiotte and
  Lefebvre}]{Blondel2008FastNetworks}
\bibinfo{author}{Blondel, V.D.}, \bibinfo{author}{Guillaume, J.L.},
  \bibinfo{author}{Lambiotte, R.}, \bibinfo{author}{Lefebvre, E.},
  \bibinfo{year}{2008}.
\newblock \bibinfo{title}{{Fast unfolding of communities in large networks}}.
\newblock \bibinfo{journal}{Journal of Statistical Mechanics: Theory and
  Experiment} \bibinfo{volume}{2008}, \bibinfo{pages}{P10008}.
\bibitem[{Buettner et~al.(2015)Buettner, Natarajan, Casale, Proserpio,
  Scialdone, Theis, Teichmann, Marioni and
  Stegle}]{Buettner2015ComputationalCells}
\bibinfo{author}{Buettner, F.}, \bibinfo{author}{Natarajan, K.N.},
  \bibinfo{author}{Casale, F.P.}, \bibinfo{author}{Proserpio, V.},
  \bibinfo{author}{Scialdone, A.}, \bibinfo{author}{Theis, F.J.},
  \bibinfo{author}{Teichmann, S.A.}, \bibinfo{author}{Marioni, J.C.},
  \bibinfo{author}{Stegle, O.}, \bibinfo{year}{2015}.
\newblock \bibinfo{title}{{Computational analysis of cell-to-cell heterogeneity
  in single-cell RNA-sequencing data reveals hidden subpopulations of cells}}.
\newblock \bibinfo{journal}{Nature Biotechnology} \bibinfo{volume}{33},
  \bibinfo{pages}{155--160}.
\bibitem[{Butler et~al.(2018)Butler, Hoffman, Smibert, Papalexi and
  Satija}]{Butler2018IntegratingSpecies}
\bibinfo{author}{Butler, A.}, \bibinfo{author}{Hoffman, P.},
  \bibinfo{author}{Smibert, P.}, \bibinfo{author}{Papalexi, E.},
  \bibinfo{author}{Satija, R.}, \bibinfo{year}{2018}.
\newblock \bibinfo{title}{{Integrating single-cell transcriptomic data across
  different conditions, technologies, and species}}.
\newblock \bibinfo{journal}{Nature Biotechnology} \bibinfo{volume}{36},
  \bibinfo{pages}{411--420}.
\bibitem[{Cole et~al.(2019)Cole, Risso, Wagner, DeTomaso, Ngai, Purdom, Dudoit
  and Yosef}]{Cole2019PerformanceRNA-Seq}
\bibinfo{author}{Cole, M.B.}, \bibinfo{author}{Risso, D.},
  \bibinfo{author}{Wagner, A.}, \bibinfo{author}{DeTomaso, D.},
  \bibinfo{author}{Ngai, J.}, \bibinfo{author}{Purdom, E.},
  \bibinfo{author}{Dudoit, S.}, \bibinfo{author}{Yosef, N.},
  \bibinfo{year}{2019}.
\newblock \bibinfo{title}{{Performance Assessment and Selection of
  Normalization Procedures for Single-Cell RNA-Seq}}.
\newblock \bibinfo{journal}{Cell Systems} \bibinfo{volume}{8},
  \bibinfo{pages}{315--328}.
\bibitem[{Crow and Gillis(2018)}]{Crow2018Co-expressionSin}
\bibinfo{author}{Crow, M.}, \bibinfo{author}{Gillis, J.}, \bibinfo{year}{2018}.
\newblock \bibinfo{title}{{Co-expression in Single-Cell Analysis: Saving Grace
  or Original Sin?}}
\newblock \bibinfo{journal}{Trends in genetics : TIG} \bibinfo{volume}{34},
  \bibinfo{pages}{823--831}.
\bibitem[{Davie et~al.(2018)Davie, Janssens, Koldere, De~Waegeneer, Pech,
  Kreft, Aibar, Makhzami, Christiaens, Bravo Gonzalez-Blas, Poovathingal,
  Hulselmans, Spanier, Moerman, Vanspauwen, Geurs, Voet, Lammertyn, Thienpont,
  Liu, Konstantinides, Fiers, Verstreken and Aerts}]{Davie2018ABrain}
\bibinfo{author}{Davie, K.}, \bibinfo{author}{Janssens, J.},
  \bibinfo{author}{Koldere, D.}, \bibinfo{author}{De~Waegeneer, M.},
  \bibinfo{author}{Pech, U.}, \bibinfo{author}{Kreft, A.},
  \bibinfo{author}{Aibar, S.}, \bibinfo{author}{Makhzami, S.},
  \bibinfo{author}{Christiaens, V.}, \bibinfo{author}{Bravo Gonzalez-Blas, C.},
  \bibinfo{author}{Poovathingal, S.}, \bibinfo{author}{Hulselmans, G.},
  \bibinfo{author}{Spanier, K.I.}, \bibinfo{author}{Moerman, T.},
  \bibinfo{author}{Vanspauwen, B.}, \bibinfo{author}{Geurs, S.},
  \bibinfo{author}{Voet, T.}, \bibinfo{author}{Lammertyn, J.},
  \bibinfo{author}{Thienpont, B.}, \bibinfo{author}{Liu, S.},
  \bibinfo{author}{Konstantinides, N.}, \bibinfo{author}{Fiers, M.},
  \bibinfo{author}{Verstreken, P.}, \bibinfo{author}{Aerts, S.},
  \bibinfo{year}{2018}.
\newblock \bibinfo{title}{{A Single-Cell Transcriptome Atlas of the Aging
  Drosophila Brain.}}
\newblock \bibinfo{journal}{Cell} \bibinfo{volume}{174},
  \bibinfo{pages}{982--998}.
\bibitem[{Ding et~al.(2015)Ding, Zheng, Zhu, Li, Jia, Ai, Wildberg and
  Wang}]{Ding2015NormalizationExperiments}
\bibinfo{author}{Ding, B.}, \bibinfo{author}{Zheng, L.}, \bibinfo{author}{Zhu,
  Y.}, \bibinfo{author}{Li, N.}, \bibinfo{author}{Jia, H.},
  \bibinfo{author}{Ai, R.}, \bibinfo{author}{Wildberg, A.},
  \bibinfo{author}{Wang, W.}, \bibinfo{year}{2015}.
\newblock \bibinfo{title}{{Normalization and noise reduction for single cell
  RNA-seq experiments}}.
\newblock \bibinfo{journal}{Bioinformatics} \bibinfo{volume}{31},
  \bibinfo{pages}{2225--2227}.
\bibitem[{Ding et~al.(2018)Ding, Condon and Shah}]{Ding2018InterpretableModels}
\bibinfo{author}{Ding, J.}, \bibinfo{author}{Condon, A.},
  \bibinfo{author}{Shah, S.P.}, \bibinfo{year}{2018}.
\newblock \bibinfo{title}{{Interpretable dimensionality reduction of single
  cell transcriptome data with deep generative models}}.
\newblock \bibinfo{journal}{Nature Communications} \bibinfo{volume}{9},
  \bibinfo{pages}{2002}.
\bibitem[{Du{\`{o}} et~al.(2018)Du{\`{o}}, Robinson and Soneson}]{Duo2018AData}
\bibinfo{author}{Du{\`{o}}, A.}, \bibinfo{author}{Robinson, M.D.},
  \bibinfo{author}{Soneson, C.}, \bibinfo{year}{2018}.
\newblock \bibinfo{title}{{A systematic performance evaluation of clustering
  methods for single-cell RNA-seq data}}.
\newblock \bibinfo{journal}{F1000Research} \bibinfo{volume}{7},
  \bibinfo{pages}{1141}.
\bibitem[{Ester et~al.(1996)Ester, Kriegel, Sander and Xu}]{Ester1996}
\bibinfo{author}{Ester, M.}, \bibinfo{author}{Kriegel, H.P.},
  \bibinfo{author}{Sander, J.}, \bibinfo{author}{Xu, X.}, \bibinfo{year}{1996}.
\newblock \bibinfo{title}{A density-based algorithm for discovering clusters a
  density-based algorithm for discovering clusters in large spatial databases
  with noise}, in: \bibinfo{booktitle}{Proceedings of the Second International
  Conference on Knowledge Discovery and Data Mining}, \bibinfo{publisher}{AAAI
  Press}. pp. \bibinfo{pages}{226--231}.
\bibitem[{Finak et~al.(2015)Finak, McDavid, Yajima, Deng, Gersuk, Shalek,
  Slichter, Miller, McElrath, Prlic, Linsley and Gottardo}]{Finak2015MAST:Data}
\bibinfo{author}{Finak, G.}, \bibinfo{author}{McDavid, A.},
  \bibinfo{author}{Yajima, M.}, \bibinfo{author}{Deng, J.},
  \bibinfo{author}{Gersuk, V.}, \bibinfo{author}{Shalek, A.K.},
  \bibinfo{author}{Slichter, C.K.}, \bibinfo{author}{Miller, H.W.},
  \bibinfo{author}{McElrath, M.J.}, \bibinfo{author}{Prlic, M.},
  \bibinfo{author}{Linsley, P.S.}, \bibinfo{author}{Gottardo, R.},
  \bibinfo{year}{2015}.
\newblock \bibinfo{title}{{MAST: a flexible statistical framework for assessing
  transcriptional changes and characterizing heterogeneity in single-cell RNA
  sequencing data}}.
\newblock \bibinfo{journal}{Genome Biology} \bibinfo{volume}{16},
  \bibinfo{pages}{278}.
\bibitem[{Freytag et~al.(2018)Freytag, Tian, L{\"{o}}nnstedt, Ng and
  Bahlo}]{Freytag2018ComparisonData}
\bibinfo{author}{Freytag, S.}, \bibinfo{author}{Tian, L.},
  \bibinfo{author}{L{\"{o}}nnstedt, I.}, \bibinfo{author}{Ng, M.},
  \bibinfo{author}{Bahlo, M.}, \bibinfo{year}{2018}.
\newblock \bibinfo{title}{{Comparison of clustering tools in R for medium-sized
  10x Genomics single-cell RNA-sequencing data}}.
\newblock \bibinfo{journal}{F1000Research} \bibinfo{volume}{7},
  \bibinfo{pages}{1297}.
\bibitem[{Gr{\"{u}}n et~al.(2015)Gr{\"{u}}n, Lyubimova, Kester, Wiebrands,
  Basak, Sasaki, Clevers and van Oudenaarden}]{Grun2015Single-cellTypes}
\bibinfo{author}{Gr{\"{u}}n, D.}, \bibinfo{author}{Lyubimova, A.},
  \bibinfo{author}{Kester, L.}, \bibinfo{author}{Wiebrands, K.},
  \bibinfo{author}{Basak, O.}, \bibinfo{author}{Sasaki, N.},
  \bibinfo{author}{Clevers, H.}, \bibinfo{author}{van Oudenaarden, A.},
  \bibinfo{year}{2015}.
\newblock \bibinfo{title}{{Single-cell messenger RNA sequencing reveals rare
  intestinal cell types}}.
\newblock \bibinfo{journal}{Nature} \bibinfo{volume}{525},
  \bibinfo{pages}{251--255}.
\bibitem[{Gr{\"{u}}n et~al.(2016)Gr{\"{u}}n, Muraro, Boisset, Wiebrands,
  Lyubimova, Dharmadhikari, van~den Born, van Es, Jansen, Clevers, de~Koning
  and van Oudenaarden}]{Grun2016DeData}
\bibinfo{author}{Gr{\"{u}}n, D.}, \bibinfo{author}{Muraro, M.J.},
  \bibinfo{author}{Boisset, J.C.}, \bibinfo{author}{Wiebrands, K.},
  \bibinfo{author}{Lyubimova, A.}, \bibinfo{author}{Dharmadhikari, G.},
  \bibinfo{author}{van~den Born, M.}, \bibinfo{author}{van Es, J.},
  \bibinfo{author}{Jansen, E.}, \bibinfo{author}{Clevers, H.},
  \bibinfo{author}{de~Koning, E.J.}, \bibinfo{author}{van Oudenaarden, A.},
  \bibinfo{year}{2016}.
\newblock \bibinfo{title}{{De Novo Prediction of Stem Cell Identity using
  Single-Cell Transcriptome Data}}.
\newblock \bibinfo{journal}{Cell Stem Cell} \bibinfo{volume}{19},
  \bibinfo{pages}{266--277}.
\bibitem[{Gr{\"{u}}n and van Oudenaarden(2015)}]{Grun2015DesignExperiments.}
\bibinfo{author}{Gr{\"{u}}n, D.}, \bibinfo{author}{van Oudenaarden, A.},
  \bibinfo{year}{2015}.
\newblock \bibinfo{title}{{Design and Analysis of Single-Cell Sequencing
  Experiments.}}
\newblock \bibinfo{journal}{Cell} \bibinfo{volume}{163},
  \bibinfo{pages}{799--810}.
\bibitem[{Guo et~al.(2015)Guo, Wang, Potter, Whitsett and Xu}]{Guo2015SINCERA}
\bibinfo{author}{Guo, M.}, \bibinfo{author}{Wang, H.}, \bibinfo{author}{Potter,
  S.S.}, \bibinfo{author}{Whitsett, J.A.}, \bibinfo{author}{Xu, Y.},
  \bibinfo{year}{2015}.
\newblock \bibinfo{title}{{SINCERA: A Pipeline for Single-Cell RNA-Seq
  Profiling Analysis}}.
\newblock \bibinfo{journal}{PLOS Computational Biology} \bibinfo{volume}{11},
  \bibinfo{pages}{e1004575}.
\bibitem[{Haghverdi et~al.(2018)Haghverdi, Lun, Morgan and
  Marioni}]{Haghverdi2018BatchNeighbors}
\bibinfo{author}{Haghverdi, L.}, \bibinfo{author}{Lun, A.T.L.},
  \bibinfo{author}{Morgan, M.D.}, \bibinfo{author}{Marioni, J.C.},
  \bibinfo{year}{2018}.
\newblock \bibinfo{title}{{Batch effects in single-cell RNA-sequencing data are
  corrected by matching mutual nearest neighbors}}.
\newblock \bibinfo{journal}{Nature Biotechnology} \bibinfo{volume}{36},
  \bibinfo{pages}{421--427}.
\bibitem[{Han et~al.(2018)Han, Wang, Zhou, Fei, Sun, Lai, Saadatpour, Zhou,
  Chen, Ye, Huang, Xu, Huang, Jiang, Jiang, Mao, Chen, Lu, Xie, Fang, Wang,
  Yue, Li, Huang, Orkin, Yuan, Chen and Guo}]{Han2018MappingMicrowell-Seq}
\bibinfo{author}{Han, X.}, \bibinfo{author}{Wang, R.}, \bibinfo{author}{Zhou,
  Y.}, \bibinfo{author}{Fei, L.}, \bibinfo{author}{Sun, H.},
  \bibinfo{author}{Lai, S.}, \bibinfo{author}{Saadatpour, A.},
  \bibinfo{author}{Zhou, Z.}, \bibinfo{author}{Chen, H.}, \bibinfo{author}{Ye,
  F.}, \bibinfo{author}{Huang, D.}, \bibinfo{author}{Xu, Y.},
  \bibinfo{author}{Huang, W.}, \bibinfo{author}{Jiang, M.},
  \bibinfo{author}{Jiang, X.}, \bibinfo{author}{Mao, J.},
  \bibinfo{author}{Chen, Y.}, \bibinfo{author}{Lu, C.}, \bibinfo{author}{Xie,
  J.}, \bibinfo{author}{Fang, Q.}, \bibinfo{author}{Wang, Y.},
  \bibinfo{author}{Yue, R.}, \bibinfo{author}{Li, T.}, \bibinfo{author}{Huang,
  H.}, \bibinfo{author}{Orkin, S.H.}, \bibinfo{author}{Yuan, G.C.},
  \bibinfo{author}{Chen, M.}, \bibinfo{author}{Guo, G.}, \bibinfo{year}{2018}.
\newblock \bibinfo{title}{{Mapping the Mouse Cell Atlas by Microwell-Seq}}.
\newblock \bibinfo{journal}{Cell} \bibinfo{volume}{172},
  \bibinfo{pages}{1091--1107}.
\bibitem[{Herman et~al.(2018)Herman, {Sagar} and
  Gr{\"{u}}n}]{Herman2018FateIDData}
\bibinfo{author}{Herman, J.S.}, \bibinfo{author}{{Sagar}},
  \bibinfo{author}{Gr{\"{u}}n, D.}, \bibinfo{year}{2018}.
\newblock \bibinfo{title}{{FateID infers cell fate bias in multipotent
  progenitors from single-cell RNA-seq data}}.
\newblock \bibinfo{journal}{Nature Methods} \bibinfo{volume}{15},
  \bibinfo{pages}{379--386}.
\bibitem[{Hubert and Arabie(1985)}]{Hubert1985ComparingPartitions}
\bibinfo{author}{Hubert, L.}, \bibinfo{author}{Arabie, P.},
  \bibinfo{year}{1985}.
\newblock \bibinfo{title}{{Comparing partitions}}.
\newblock \bibinfo{journal}{Journal of Classification} \bibinfo{volume}{2},
  \bibinfo{pages}{193--218}.
\bibitem[{Jiang et~al.(2016a)Jiang, Chen, Pinello and
  Yuan}]{Jiang2016GiniClust:Index}
\bibinfo{author}{Jiang, L.}, \bibinfo{author}{Chen, H.},
  \bibinfo{author}{Pinello, L.}, \bibinfo{author}{Yuan, G.C.},
  \bibinfo{year}{2016}a.
\newblock \bibinfo{title}{{GiniClust: detecting rare cell types from
  single-cell gene expression data with Gini index}}.
\newblock \bibinfo{journal}{Genome Biology} \bibinfo{volume}{17},
  \bibinfo{pages}{144}.
\bibitem[{Jiang et~al.(2016b)Jiang, Thomson and
  Stewart}]{Jiang2016QualitySinQC}
\bibinfo{author}{Jiang, P.}, \bibinfo{author}{Thomson, J.A.},
  \bibinfo{author}{Stewart, R.}, \bibinfo{year}{2016}b.
\newblock \bibinfo{title}{{Quality control of single-cell RNA-seq by SinQC}}.
\newblock \bibinfo{journal}{Bioinformatics} \bibinfo{volume}{32},
  \bibinfo{pages}{2514--2516}.
\bibitem[{Kharchenko et~al.(2014)Kharchenko, Silberstein and
  Scadden}]{Kharchenko2014BayesianAnalysis}
\bibinfo{author}{Kharchenko, P.V.}, \bibinfo{author}{Silberstein, L.},
  \bibinfo{author}{Scadden, D.T.}, \bibinfo{year}{2014}.
\newblock \bibinfo{title}{{Bayesian approach to single-cell differential
  expression analysis}}.
\newblock \bibinfo{journal}{Nature Methods} \bibinfo{volume}{11},
  \bibinfo{pages}{740--742}.
\bibitem[{Kiselev et~al.(2019)Kiselev, Andrews and
  Hemberg}]{Kiselev2019ChallengesData}
\bibinfo{author}{Kiselev, V.Y.}, \bibinfo{author}{Andrews, T.S.},
  \bibinfo{author}{Hemberg, M.}, \bibinfo{year}{2019}.
\newblock \bibinfo{title}{{Challenges in unsupervised clustering of single-cell
  RNA-seq data}}.
\newblock \bibinfo{journal}{Nature Reviews Genetics} \bibinfo{volume}{20},
  \bibinfo{pages}{273--282}.
\bibitem[{Kiselev et~al.(2017)Kiselev, Kirschner, Schaub, Andrews, Yiu,
  Chandra, Natarajan, Reik, Barahona, Green and Hemberg}]{Kiselev2017SC3:Data}
\bibinfo{author}{Kiselev, V.Y.}, \bibinfo{author}{Kirschner, K.},
  \bibinfo{author}{Schaub, M.T.}, \bibinfo{author}{Andrews, T.},
  \bibinfo{author}{Yiu, A.}, \bibinfo{author}{Chandra, T.},
  \bibinfo{author}{Natarajan, K.N.}, \bibinfo{author}{Reik, W.},
  \bibinfo{author}{Barahona, M.}, \bibinfo{author}{Green, A.R.},
  \bibinfo{author}{Hemberg, M.}, \bibinfo{year}{2017}.
\newblock \bibinfo{title}{{SC3: consensus clustering of single-cell RNA-seq
  data}}.
\newblock \bibinfo{journal}{Nature Methods} \bibinfo{volume}{14},
  \bibinfo{pages}{483--486}.
\bibitem[{Klein et~al.(2015)Klein, Mazutis, Akartuna, Tallapragada, Veres, Li,
  Peshkin, Weitz and Kirschner}]{Klein2015DropletCells}
\bibinfo{author}{Klein, A.M.}, \bibinfo{author}{Mazutis, L.},
  \bibinfo{author}{Akartuna, I.}, \bibinfo{author}{Tallapragada, N.},
  \bibinfo{author}{Veres, A.}, \bibinfo{author}{Li, V.},
  \bibinfo{author}{Peshkin, L.}, \bibinfo{author}{Weitz, D.A.},
  \bibinfo{author}{Kirschner, M.W.}, \bibinfo{year}{2015}.
\newblock \bibinfo{title}{{Droplet barcoding for single-cell transcriptomics
  applied to embryonic stem cells}}.
\newblock \bibinfo{journal}{Cell} \bibinfo{volume}{161},
  \bibinfo{pages}{1187--1201}.
\bibitem[{Li et~al.(2017)Li, Courtois, Sengupta, Tan, Chen, Goh, Kong, Chua,
  Hon, Tan, Wong, Choi, Wee, Hillmer, Tan, Robson and
  Prabhakar}]{Li2017ReferenceTumors}
\bibinfo{author}{Li, H.}, \bibinfo{author}{Courtois, E.T.},
  \bibinfo{author}{Sengupta, D.}, \bibinfo{author}{Tan, Y.},
  \bibinfo{author}{Chen, K.H.}, \bibinfo{author}{Goh, J.J.L.},
  \bibinfo{author}{Kong, S.L.}, \bibinfo{author}{Chua, C.},
  \bibinfo{author}{Hon, L.K.}, \bibinfo{author}{Tan, W.S.},
  \bibinfo{author}{Wong, M.}, \bibinfo{author}{Choi, P.J.},
  \bibinfo{author}{Wee, L.J.K.}, \bibinfo{author}{Hillmer, A.M.},
  \bibinfo{author}{Tan, I.B.}, \bibinfo{author}{Robson, P.},
  \bibinfo{author}{Prabhakar, S.}, \bibinfo{year}{2017}.
\newblock \bibinfo{title}{{Reference component analysis of single-cell
  transcriptomes elucidates cellular heterogeneity in human colorectal
  tumors}}.
\newblock \bibinfo{journal}{Nature Genetics} \bibinfo{volume}{49},
  \bibinfo{pages}{708--718}.
\bibitem[{Lin et~al.(2017a)Lin, Jain, Kim and Bar-Joseph}]{Lin2017UsingData}
\bibinfo{author}{Lin, C.}, \bibinfo{author}{Jain, S.}, \bibinfo{author}{Kim,
  H.}, \bibinfo{author}{Bar-Joseph, Z.}, \bibinfo{year}{2017}a.
\newblock \bibinfo{title}{{Using neural networks for reducing the dimensions of
  single-cell RNA-Seq data}}.
\newblock \bibinfo{journal}{Nucleic Acids Research} \bibinfo{volume}{45},
  \bibinfo{pages}{e156--e156}.
\bibitem[{Lin et~al.(2017b)Lin, Troup and Ho}]{Lin2017CIDR:Data}
\bibinfo{author}{Lin, P.}, \bibinfo{author}{Troup, M.}, \bibinfo{author}{Ho,
  J.W.K.}, \bibinfo{year}{2017}b.
\newblock \bibinfo{title}{{CIDR: Ultrafast and accurate clustering through
  imputation for single-cell RNA-seq data}}.
\newblock \bibinfo{journal}{Genome Biology} \bibinfo{volume}{18},
  \bibinfo{pages}{59}.
\bibitem[{Linderman et~al.(2019)Linderman, Rachh, Hoskins, Steinerberger and
  Kluger}]{Linderman2019FastData}
\bibinfo{author}{Linderman, G.C.}, \bibinfo{author}{Rachh, M.},
  \bibinfo{author}{Hoskins, J.G.}, \bibinfo{author}{Steinerberger, S.},
  \bibinfo{author}{Kluger, Y.}, \bibinfo{year}{2019}.
\newblock \bibinfo{title}{{Fast interpolation-based t-SNE for improved
  visualization of single-cell RNA-seq data}}.
\newblock \bibinfo{journal}{Nature Methods} \bibinfo{volume}{16},
  \bibinfo{pages}{243--245}.
\bibitem[{Lloyd(1982)}]{Lloyd1982LeastPCM}
\bibinfo{author}{Lloyd, S.}, \bibinfo{year}{1982}.
\newblock \bibinfo{title}{{Least squares quantization in PCM}}.
\newblock \bibinfo{journal}{IEEE Transactions on Information Theory}
  \bibinfo{volume}{28}, \bibinfo{pages}{129--137}.
\bibitem[{Lun et~al.(2016a)Lun, Bach and Marioni}]{L.Lun2016PoolingCounts}
\bibinfo{author}{Lun, A.T.}, \bibinfo{author}{Bach, K.},
  \bibinfo{author}{Marioni, J.C.}, \bibinfo{year}{2016}a.
\newblock \bibinfo{title}{{Pooling across cells to normalize single-cell RNA
  sequencing data with many zero counts}}.
\newblock \bibinfo{journal}{Genome Biology} \bibinfo{volume}{17},
  \bibinfo{pages}{75}.
\bibitem[{Lun et~al.(2016b)Lun, McCarthy and Marioni}]{Lun2016AData}
\bibinfo{author}{Lun, A.T.}, \bibinfo{author}{McCarthy, D.J.},
  \bibinfo{author}{Marioni, J.C.}, \bibinfo{year}{2016}b.
\newblock \bibinfo{title}{{A step-by-step workflow for low-level analysis of
  single-cell RNA-seq data}}.
\newblock \bibinfo{journal}{F1000Research} \bibinfo{volume}{5},
  \bibinfo{pages}{2122}.
\bibitem[{McGinnis et~al.(2019)McGinnis, Murrow and
  Gartner}]{McGinnis2019DoubletFinder:Neighbors}
\bibinfo{author}{McGinnis, C.S.}, \bibinfo{author}{Murrow, L.M.},
  \bibinfo{author}{Gartner, Z.J.}, \bibinfo{year}{2019}.
\newblock \bibinfo{title}{{DoubletFinder: Doublet Detection in Single-Cell RNA
  Sequencing Data Using Artificial Nearest Neighbors}}.
\newblock \bibinfo{journal}{Cell Systems} \bibinfo{volume}{8},
  \bibinfo{pages}{329--337}.
\bibitem[{Mcinnes et~al.(2018)Mcinnes, Healy, Saul and
  Gro{\ss}berger}]{Mcinnes2018UMAP:Archive}
\bibinfo{author}{Mcinnes, L.}, \bibinfo{author}{Healy, J.},
  \bibinfo{author}{Saul, N.}, \bibinfo{author}{Gro{\ss}berger, L.},
  \bibinfo{year}{2018}.
\newblock \bibinfo{title}{{UMAP: Uniform Manifold Approximation and
  Projection}}.
\newblock \bibinfo{journal}{Journal of Open Source Software}
  \DOIprefix\doi{10.21105/joss.00861}.
\bibitem[{Ntranos et~al.(2016)Ntranos, Kamath, Zhang, Pachter and
  Tse}]{Ntranos2016FastCounts}
\bibinfo{author}{Ntranos, V.}, \bibinfo{author}{Kamath, G.M.},
  \bibinfo{author}{Zhang, J.M.}, \bibinfo{author}{Pachter, L.},
  \bibinfo{author}{Tse, D.N.}, \bibinfo{year}{2016}.
\newblock \bibinfo{title}{{Fast and accurate single-cell RNA-seq analysis by
  clustering of transcript-compatibility counts}}.
\newblock \bibinfo{journal}{Genome Biology} \bibinfo{volume}{17},
  \bibinfo{pages}{112}.
\bibitem[{Patel et~al.(2014)Patel, Tirosh, Trombetta, Shalek, Gillespie,
  Wakimoto, Cahill, Nahed, Curry, Martuza, Louis, Rozenblatt-Rosen, Suv{\`{a}},
  Regev and Bernstein}]{Patel2014Single-cellGlioblastoma}
\bibinfo{author}{Patel, A.P.}, \bibinfo{author}{Tirosh, I.},
  \bibinfo{author}{Trombetta, J.J.}, \bibinfo{author}{Shalek, A.K.},
  \bibinfo{author}{Gillespie, S.M.}, \bibinfo{author}{Wakimoto, H.},
  \bibinfo{author}{Cahill, D.P.}, \bibinfo{author}{Nahed, B.V.},
  \bibinfo{author}{Curry, W.T.}, \bibinfo{author}{Martuza, R.L.},
  \bibinfo{author}{Louis, D.N.}, \bibinfo{author}{Rozenblatt-Rosen, O.},
  \bibinfo{author}{Suv{\`{a}}, M.L.}, \bibinfo{author}{Regev, A.},
  \bibinfo{author}{Bernstein, B.E.}, \bibinfo{year}{2014}.
\newblock \bibinfo{title}{{Single-cell RNA-seq highlights intratumoral
  heterogeneity in primary glioblastoma.}}
\newblock \bibinfo{journal}{Science (New York, N.Y.)} \bibinfo{volume}{344},
  \bibinfo{pages}{1396--401}.
\bibitem[{Prabhakaran et~al.(2016)Prabhakaran, Azizi, Carr and
  PeÃ¢er}]{Prabhakaran16}
\bibinfo{author}{Prabhakaran, S.}, \bibinfo{author}{Azizi, E.},
  \bibinfo{author}{Carr, A.}, \bibinfo{author}{PeÃ¢er, D.},
  \bibinfo{year}{2016}.
\newblock \bibinfo{title}{Dirichlet process mixture model for correcting
  technical variation in single-cell gene expression data}, in:
  \bibinfo{editor}{Balcan, M.F.}, \bibinfo{editor}{Weinberger, K.Q.} (Eds.),
  \bibinfo{booktitle}{Proceedings of The 33rd International Conference on
  Machine Learning}, \bibinfo{publisher}{PMLR}, \bibinfo{address}{New York, New
  York, USA}. pp. \bibinfo{pages}{1070--1079}.
\bibitem[{Qiu et~al.(2017a)Qiu, Hill, Packer, Lin, Ma and
  Trapnell}]{Qiu2017Single-cellCensus}
\bibinfo{author}{Qiu, X.}, \bibinfo{author}{Hill, A.}, \bibinfo{author}{Packer,
  J.}, \bibinfo{author}{Lin, D.}, \bibinfo{author}{Ma, Y.A.},
  \bibinfo{author}{Trapnell, C.}, \bibinfo{year}{2017}a.
\newblock \bibinfo{title}{{Single-cell mRNA quantification and differential
  analysis with Census}}.
\newblock \bibinfo{journal}{Nature Methods} \bibinfo{volume}{14},
  \bibinfo{pages}{309--315}.
\bibitem[{Qiu et~al.(2017b)Qiu, Mao, Tang, Wang, Chawla, Pliner and
  Trapnell}]{Qiu2017ReversedTrajectories}
\bibinfo{author}{Qiu, X.}, \bibinfo{author}{Mao, Q.}, \bibinfo{author}{Tang,
  Y.}, \bibinfo{author}{Wang, L.}, \bibinfo{author}{Chawla, R.},
  \bibinfo{author}{Pliner, H.A.}, \bibinfo{author}{Trapnell, C.},
  \bibinfo{year}{2017}b.
\newblock \bibinfo{title}{{Reversed graph embedding resolves complex
  single-cell trajectories}}.
\newblock \bibinfo{journal}{Nature Methods} \bibinfo{volume}{14},
  \bibinfo{pages}{979--982}.
\bibitem[{Risso et~al.(2014)Risso, Ngai, Speed and
  Dudoit}]{Risso2014NormalizationSamples}
\bibinfo{author}{Risso, D.}, \bibinfo{author}{Ngai, J.},
  \bibinfo{author}{Speed, T.P.}, \bibinfo{author}{Dudoit, S.},
  \bibinfo{year}{2014}.
\newblock \bibinfo{title}{{Normalization of RNA-seq data using factor analysis
  of control genes or samples}}.
\newblock \bibinfo{journal}{Nature Biotechnology} \bibinfo{volume}{32},
  \bibinfo{pages}{896--902}.
\bibitem[{Rosenberg and Hirschberg(2007)}]{Rosenberg2007V-Measure:Measure}
\bibinfo{author}{Rosenberg, A.}, \bibinfo{author}{Hirschberg, J.},
  \bibinfo{year}{2007}.
\newblock \bibinfo{title}{{V-Measure: A conditional entropy-based external
  cluster evaluation measure}}.
\newblock \bibinfo{journal}{Proceedings of the 2007 joint conference on
  empirical methods in natural language processing and computational natural
  language learning (EMNLP-CoNLL)} , \bibinfo{pages}{410--420}.
\bibitem[{Rozenblatt-Rosen et~al.(2017)Rozenblatt-Rosen, Stubbington, Regev and
  Teichmann}]{Rozenblatt-Rosen2017TheReality}
\bibinfo{author}{Rozenblatt-Rosen, O.}, \bibinfo{author}{Stubbington, M.J.T.},
  \bibinfo{author}{Regev, A.}, \bibinfo{author}{Teichmann, S.A.},
  \bibinfo{year}{2017}.
\newblock \bibinfo{title}{{The Human Cell Atlas: from vision to reality}}.
\newblock \bibinfo{journal}{Nature} \bibinfo{volume}{550},
  \bibinfo{pages}{451--453}.
\bibitem[{Satija et~al.(2015)Satija, Farrell, Gennert, Schier and
  Regev}]{Satija2015SpatialData}
\bibinfo{author}{Satija, R.}, \bibinfo{author}{Farrell, J.A.},
  \bibinfo{author}{Gennert, D.}, \bibinfo{author}{Schier, A.F.},
  \bibinfo{author}{Regev, A.}, \bibinfo{year}{2015}.
\newblock \bibinfo{title}{{Spatial reconstruction of single-cell gene
  expression data}}.
\newblock \bibinfo{journal}{Nature Biotechnology} \bibinfo{volume}{33},
  \bibinfo{pages}{495--502}.
\bibitem[{Shalek et~al.(2014)Shalek, Satija, Shuga, Trombetta, Gennert, Lu,
  Chen, Gertner, Gaublomme, Yosef, Schwartz, Fowler, Weaver, Wang, Wang, Ding,
  Raychowdhury, Friedman, Hacohen, Park, May and
  Regev}]{Shalek2014Single-cellVariation}
\bibinfo{author}{Shalek, A.K.}, \bibinfo{author}{Satija, R.},
  \bibinfo{author}{Shuga, J.}, \bibinfo{author}{Trombetta, J.J.},
  \bibinfo{author}{Gennert, D.}, \bibinfo{author}{Lu, D.},
  \bibinfo{author}{Chen, P.}, \bibinfo{author}{Gertner, R.S.},
  \bibinfo{author}{Gaublomme, J.T.}, \bibinfo{author}{Yosef, N.},
  \bibinfo{author}{Schwartz, S.}, \bibinfo{author}{Fowler, B.},
  \bibinfo{author}{Weaver, S.}, \bibinfo{author}{Wang, J.},
  \bibinfo{author}{Wang, X.}, \bibinfo{author}{Ding, R.},
  \bibinfo{author}{Raychowdhury, R.}, \bibinfo{author}{Friedman, N.},
  \bibinfo{author}{Hacohen, N.}, \bibinfo{author}{Park, H.},
  \bibinfo{author}{May, A.P.}, \bibinfo{author}{Regev, A.},
  \bibinfo{year}{2014}.
\newblock \bibinfo{title}{{Single-cell RNA-seq reveals dynamic paracrine
  control of cellular variation}}.
\newblock \bibinfo{journal}{Nature} \bibinfo{volume}{510},
  \bibinfo{pages}{363--369}.
\bibitem[{Shapiro et~al.(2013)Shapiro, Biezuner and
  Linnarsson}]{Shapiro2013Single-cellScience}
\bibinfo{author}{Shapiro, E.}, \bibinfo{author}{Biezuner, T.},
  \bibinfo{author}{Linnarsson, S.}, \bibinfo{year}{2013}.
\newblock \bibinfo{title}{{Single-cell sequencing-based technologies will
  revolutionize whole-organism science}}.
\newblock \bibinfo{journal}{Nature Reviews Genetics} \bibinfo{volume}{14},
  \bibinfo{pages}{618--630}.
\bibitem[{Tibshirani et~al.(2001)Tibshirani, Walther and
  Hastie}]{Tibshirani2001EstimatingStatistic}
\bibinfo{author}{Tibshirani, R.}, \bibinfo{author}{Walther, G.},
  \bibinfo{author}{Hastie, T.}, \bibinfo{year}{2001}.
\newblock \bibinfo{title}{{Estimating the number of clusters in a data set via
  the gap statistic}}.
\newblock \bibinfo{journal}{Journal of the Royal Statistical Society: Series B
  (Statistical Methodology)} \bibinfo{volume}{63}, \bibinfo{pages}{411--423}.
\bibitem[{Tsafrir et~al.(2005)Tsafrir, Tsafrir, Ein-Dor, Zuk, Notterman and
  Domany}]{Tsafrir2005SortingMatrices}
\bibinfo{author}{Tsafrir, D.}, \bibinfo{author}{Tsafrir, I.},
  \bibinfo{author}{Ein-Dor, L.}, \bibinfo{author}{Zuk, O.},
  \bibinfo{author}{Notterman, D.}, \bibinfo{author}{Domany, E.},
  \bibinfo{year}{2005}.
\newblock \bibinfo{title}{{Sorting points into neighborhoods (SPIN): data
  analysis and visualization by ordering distance matrices}}.
\newblock \bibinfo{journal}{Bioinformatics} \bibinfo{volume}{21},
  \bibinfo{pages}{2301--2308}.
\bibitem[{van Unen et~al.(2017)van Unen, H{\"{o}}llt, Pezzotti, Li, Reinders,
  Eisemann, Koning, Vilanova and Lelieveldt}]{vanUnen2017VisualTypes}
\bibinfo{author}{van Unen, V.}, \bibinfo{author}{H{\"{o}}llt, T.},
  \bibinfo{author}{Pezzotti, N.}, \bibinfo{author}{Li, N.},
  \bibinfo{author}{Reinders, M.J.T.}, \bibinfo{author}{Eisemann, E.},
  \bibinfo{author}{Koning, F.}, \bibinfo{author}{Vilanova, A.},
  \bibinfo{author}{Lelieveldt, B.P.F.}, \bibinfo{year}{2017}.
\newblock \bibinfo{title}{{Visual analysis of mass cytometry data by
  hierarchical stochastic neighbour embedding reveals rare cell types}}.
\newblock \bibinfo{journal}{Nature Communications} \bibinfo{volume}{8},
  \bibinfo{pages}{1740}.
\bibitem[{Usoskin et~al.(2015)Usoskin, Furlan, Islam, Abdo, L{\"{o}}nnerberg,
  Lou, Hjerling-Leffler, Haeggstr{\"{o}}m, Kharchenko, Kharchenko, Linnarsson
  and Ernfors}]{Usoskin2015UnbiasedSequencing}
\bibinfo{author}{Usoskin, D.}, \bibinfo{author}{Furlan, A.},
  \bibinfo{author}{Islam, S.}, \bibinfo{author}{Abdo, H.},
  \bibinfo{author}{L{\"{o}}nnerberg, P.}, \bibinfo{author}{Lou, D.},
  \bibinfo{author}{Hjerling-Leffler, J.}, \bibinfo{author}{Haeggstr{\"{o}}m,
  J.}, \bibinfo{author}{Kharchenko, O.}, \bibinfo{author}{Kharchenko, P.V.},
  \bibinfo{author}{Linnarsson, S.}, \bibinfo{author}{Ernfors, P.},
  \bibinfo{year}{2015}.
\newblock \bibinfo{title}{{Unbiased classification of sensory neuron types by
  large-scale single-cell RNA sequencing}}.
\newblock \bibinfo{journal}{Nature Neuroscience} \bibinfo{volume}{18},
  \bibinfo{pages}{145--153}.
\bibitem[{Vallejos et~al.(2015)Vallejos, Marioni and
  Richardson}]{Vallejos2015BASiCS:Data}
\bibinfo{author}{Vallejos, C.A.}, \bibinfo{author}{Marioni, J.C.},
  \bibinfo{author}{Richardson, S.}, \bibinfo{year}{2015}.
\newblock \bibinfo{title}{{BASiCS: Bayesian Analysis of Single-Cell Sequencing
  Data}}.
\newblock \bibinfo{journal}{PLOS Computational Biology} \bibinfo{volume}{11},
  \bibinfo{pages}{e1004333}.
\bibitem[{Vallejos et~al.(2017)Vallejos, Risso, Scialdone, Dudoit and
  Marioni}]{Vallejos2017NormalizingOpportunities}
\bibinfo{author}{Vallejos, C.A.}, \bibinfo{author}{Risso, D.},
  \bibinfo{author}{Scialdone, A.}, \bibinfo{author}{Dudoit, S.},
  \bibinfo{author}{Marioni, J.C.}, \bibinfo{year}{2017}.
\newblock \bibinfo{title}{{Normalizing single-cell RNA sequencing data:
  challenges and opportunities}}.
\newblock \bibinfo{journal}{Nature Methods} \bibinfo{volume}{14},
  \bibinfo{pages}{565--571}.
\bibitem[{Wang et~al.(2017)Wang, Zhu, Pierson, Ramazzotti and
  Batzoglou}]{Wang2017VisualizationLearning}
\bibinfo{author}{Wang, B.}, \bibinfo{author}{Zhu, J.},
  \bibinfo{author}{Pierson, E.}, \bibinfo{author}{Ramazzotti, D.},
  \bibinfo{author}{Batzoglou, S.}, \bibinfo{year}{2017}.
\newblock \bibinfo{title}{{Visualization and analysis of single-cell RNA-seq
  data by kernel-based similarity learning}}.
\newblock \bibinfo{journal}{Nature Methods} \bibinfo{volume}{14},
  \bibinfo{pages}{414--416}.
\bibitem[{Wang and Gu(2018)}]{Wang2018VASC:Autoencoder}
\bibinfo{author}{Wang, D.}, \bibinfo{author}{Gu, J.}, \bibinfo{year}{2018}.
\newblock \bibinfo{title}{{VASC: Dimension Reduction and Visualization of
  Single-cell RNA-seq Data by Deep Variational Autoencoder}}.
\newblock \bibinfo{journal}{Genomics, Proteomics {\&} Bioinformatics}
  \bibinfo{volume}{16}, \bibinfo{pages}{320--331}.
\bibitem[{Wolf et~al.(2018)Wolf, Angerer and Theis}]{Wolf2018SCANPY:Analysis}
\bibinfo{author}{Wolf, F.A.}, \bibinfo{author}{Angerer, P.},
  \bibinfo{author}{Theis, F.J.}, \bibinfo{year}{2018}.
\newblock \bibinfo{title}{{SCANPY: large-scale single-cell gene expression data
  analysis}}.
\newblock \bibinfo{journal}{Genome Biology} \bibinfo{volume}{19},
  \bibinfo{pages}{15}.
\bibitem[{Wolock et~al.(2019)Wolock, Lopez and Klein}]{Wolock2019Scrublet:Data}
\bibinfo{author}{Wolock, S.L.}, \bibinfo{author}{Lopez, R.},
  \bibinfo{author}{Klein, A.M.}, \bibinfo{year}{2019}.
\newblock \bibinfo{title}{{Scrublet: Computational Identification of Cell
  Doublets in Single-Cell Transcriptomic Data}}.
\newblock \bibinfo{journal}{Cell Systems} \bibinfo{volume}{8},
  \bibinfo{pages}{281--291}.
\bibitem[{Xu and Su(2015)}]{Xu2015IdentificationMethod}
\bibinfo{author}{Xu, C.}, \bibinfo{author}{Su, Z.}, \bibinfo{year}{2015}.
\newblock \bibinfo{title}{{Identification of cell types from single-cell
  transcriptomes using a novel clustering method}}.
\newblock \bibinfo{journal}{Bioinformatics} \bibinfo{volume}{31},
  \bibinfo{pages}{1974--1980}.
\bibitem[{Yang et~al.(2017)Yang, Liu, Lu, Riggs and Wu}]{Yang2017SAIC:Data}
\bibinfo{author}{Yang, L.}, \bibinfo{author}{Liu, J.}, \bibinfo{author}{Lu,
  Q.}, \bibinfo{author}{Riggs, A.D.}, \bibinfo{author}{Wu, X.},
  \bibinfo{year}{2017}.
\newblock \bibinfo{title}{{SAIC: an iterative clustering approach for analysis
  of single cell RNA-seq data}}.
\newblock \bibinfo{journal}{BMC Genomics} \bibinfo{volume}{18},
  \bibinfo{pages}{689}.
\bibitem[{Zeisel et~al.(2015)Zeisel, Mu{\~{n}}oz-Manchado, Codeluppi,
  L{\"{o}}nnerberg, La~Manno, Jur{\'{e}}us, Marques, Munguba, He, Betsholtz,
  Rolny, Castelo-Branco, Hjerling-Leffler and
  Linnarsson}]{Zeisel2015BrainRNA-seq}
\bibinfo{author}{Zeisel, A.}, \bibinfo{author}{Mu{\~{n}}oz-Manchado, A.B.},
  \bibinfo{author}{Codeluppi, S.}, \bibinfo{author}{L{\"{o}}nnerberg, P.},
  \bibinfo{author}{La~Manno, G.}, \bibinfo{author}{Jur{\'{e}}us, A.},
  \bibinfo{author}{Marques, S.}, \bibinfo{author}{Munguba, H.},
  \bibinfo{author}{He, L.}, \bibinfo{author}{Betsholtz, C.},
  \bibinfo{author}{Rolny, C.}, \bibinfo{author}{Castelo-Branco, G.},
  \bibinfo{author}{Hjerling-Leffler, J.}, \bibinfo{author}{Linnarsson, S.},
  \bibinfo{year}{2015}.
\newblock \bibinfo{title}{{Cell types in the mouse cortex and hippocampus
  revealed by single-cell RNA-seq}}.
\newblock \bibinfo{journal}{Science} \bibinfo{volume}{347},
  \bibinfo{pages}{1138--42}.
\bibitem[{Zhang et~al.(2018)Zhang, Fan, Fan, Rosenfeld and
  Tse}]{Zhang2018AnDatasets}
\bibinfo{author}{Zhang, J.M.}, \bibinfo{author}{Fan, J.}, \bibinfo{author}{Fan,
  H.C.}, \bibinfo{author}{Rosenfeld, D.}, \bibinfo{author}{Tse, D.N.},
  \bibinfo{year}{2018}.
\newblock \bibinfo{title}{{An interpretable framework for clustering
  single-cell RNA-Seq datasets}}.
\newblock \bibinfo{journal}{BMC Bioinformatics} \bibinfo{volume}{19},
  \bibinfo{pages}{93}.
\bibitem[{Zhang et~al.(2009)Zhang, Xu, Li and Su}]{Zhang2009Genome}
\bibinfo{author}{Zhang, S.}, \bibinfo{author}{Xu, M.}, \bibinfo{author}{Li,
  S.}, \bibinfo{author}{Su, Z.}, \bibinfo{year}{2009}.
\newblock \bibinfo{title}{{Genome-wide de novo prediction of cis-regulatory
  binding sites in prokaryotes}}.
\newblock \bibinfo{journal}{Nucleic Acids Research} \bibinfo{volume}{37},
  \bibinfo{pages}{e72--e72}.
\bibitem[{Zheng et~al.(2019)Zheng, Li, Liang, Wu, Pan and
  Wang}]{Zheng2019SinNLRR}
\bibinfo{author}{Zheng, R.}, \bibinfo{author}{Li, M.}, \bibinfo{author}{Liang,
  Z.}, \bibinfo{author}{Wu, F.X.}, \bibinfo{author}{Pan, Y.},
  \bibinfo{author}{Wang, J.}, \bibinfo{year}{2019}.
\newblock \bibinfo{title}{{SinNLRR: a robust subspace clustering method for
  cell type detection by non-negative and low-rank representation}}.
\newblock \bibinfo{journal}{Bioinformatics} .
\bibitem[{zurauskiene and Yau(2016)}]{zurauskiene2016PcaReduce:Profiles}
\bibinfo{author}{zurauskiene, J.}, \bibinfo{author}{Yau, C.},
  \bibinfo{year}{2016}.
\newblock \bibinfo{title}{{pcaReduce: hierarchical clustering of single cell
  transcriptional profiles}}.
\newblock \bibinfo{journal}{BMC Bioinformatics} \bibinfo{volume}{17},
  \bibinfo{pages}{140}.

\end{thebibliography}

\end{document}